\documentclass[twocolumn,amsmath,amssymb,longbibliography]{revtex4-2}

\usepackage{graphicx}
\usepackage{dcolumn}
\usepackage{bm}
\usepackage[utf8]{inputenc}
\usepackage[T1]{fontenc}
\usepackage{mathptmx}
\usepackage{etoolbox}
\usepackage{multirow}
\usepackage{xcolor}

\definecolor{lc}{HTML}{51ABE3}
\usepackage[colorlinks,allcolors=lc]{hyperref}

\makeatletter
\def\@email#1#2{%
	\endgroup
	\patchcmd{\titleblock@produce}
	{\frontmatter@RRAPformat}
	{\frontmatter@RRAPformat{\produce@RRAP{*#1\href{mailto:#2}{#2}}}\frontmatter@RRAPformat}
	{}{}
}%
\makeatother
\begin{document}

\title{Experimental observation of anomalous supralinear response\\ of single-photon detectors}

\author{Josef Hlou\v{s}ek}
\email{hlousek@optics.upol.cz}
\author{Ivo Straka}
\author{Miroslav Je\v{z}ek}
\email{jezek@optics.upol.cz}
\affiliation{Department of Optics, Faculty of Science, Palack\'{y} University, 17.~listopadu 12, 77146 Olomouc, Czechia}

\date{\today}

\begin{abstract}
	The linearity of single-photon detectors allows accurate optical measurements at low light levels and using non-classical light in spectroscopy, biomedical imaging, optical communication, and sensing. However, in practice the response of single-photon detectors can exhibit intriguing nonlinear effects that may influence the performed measurements. Here, we demonstrate a direct single-source measurement of absolute nonlinearity of single-photon detectors with unprecedented accuracy. We discover a surprising supralinear behavior of single-photon avalanche diodes and show that it cannot be explained using known theoretical models. We also fully characterize sub- and supra-linear operation regimes of superconducting nanowire single-photon detectors and uncover the supralinearity under faint continuous illumination. The results identify new detector anomalies that supersede existing knowledge of nonlinear effects at the single-photon level.
\end{abstract}

\maketitle

\section{Introduction}

\begin{figure}[ht]
	\centering
	\includegraphics[width=0.85\linewidth]{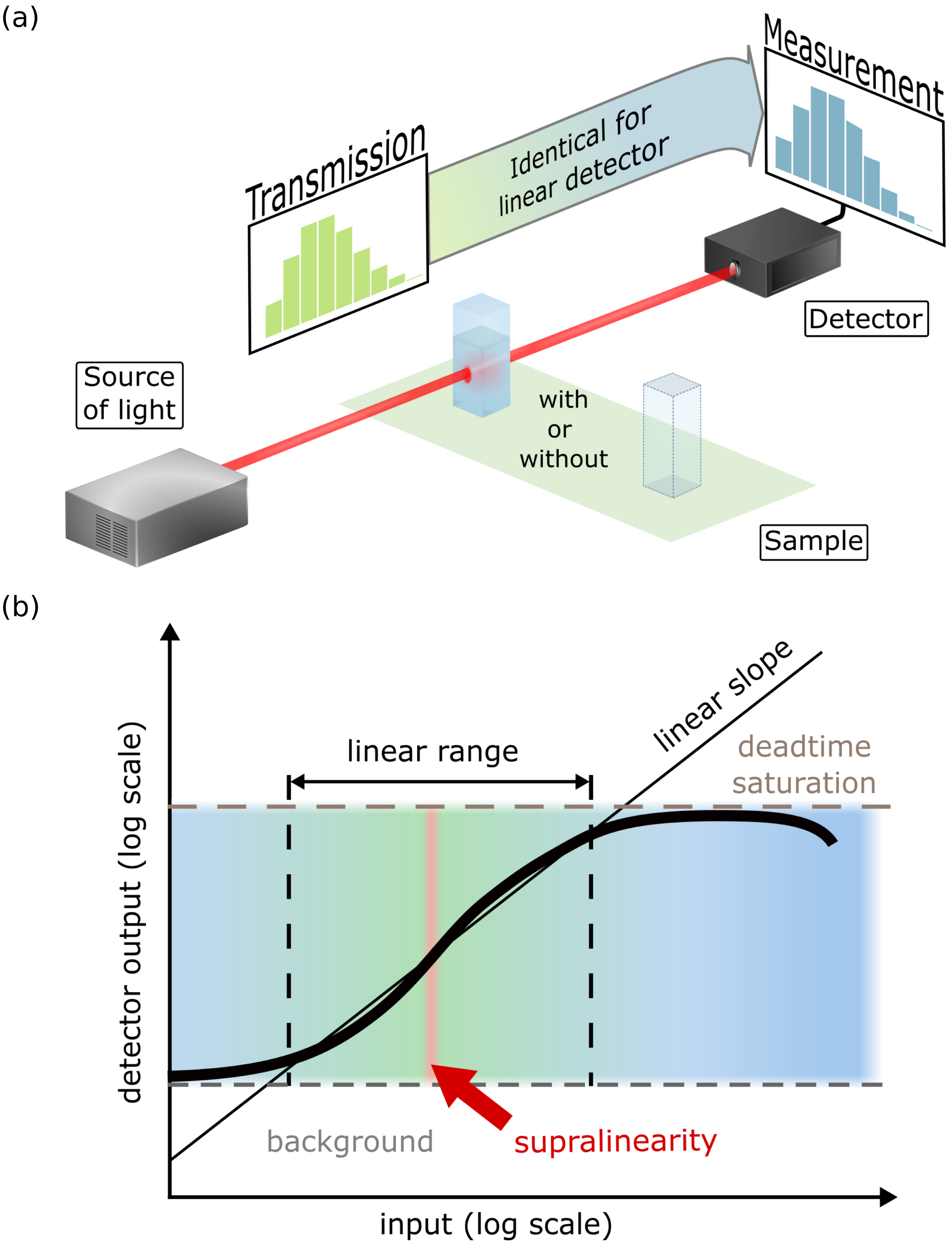}
	\caption{(A) Optical transmittance measurement of a sample relies on comparing two levels of intensity captured by a detector. Any deviation from the ideal linear detection then critically affects the measurement accuracy. Nonlinearity is a particularly vexing problem in quantum metrology when nonclassical statistics of light is often required together with SPDs. Their inherently nonlinear response distorts the photon statistics and makes it difficult to reach the quantum advantage.
		(B) Nonlinear behavior of a SPD. Apart from the background detections and saturation effects, SPDs may exhibit supralinearity.}
	\label{fig_NL}
\end{figure}

A majority of radiometric, spectroscopic, imaging, and optical communication methods rely on comparing two or more levels of light intensity measured by a photodetector and assuming that its response is proportional to the incident radiation. Transmittance measurement represents the simplest example where the optical power is detected with and without the sample under test, see Fig.~\ref{fig_NL}(A).
For an ideal detector, the power ratio would be the same as the actual transmittance of the sample.
The measurement accuracy is, however, impaired by any deviation from a perfectly linear response of the detector. 
One can correct for the detection imperfections on the condition that the model of the nonlinear effect is known and accurate enough.

With the advent of ultra-sensitive detectors and quantum-enhanced metrology, we tend to perform measurements at the ultimate sensitivity levels dictated by the laws of physics \cite{Giovannetti2011,Rohde2015}. The goal is to reach the quantum advantage regime---that is, to improve the sensitivity of a measurement beyond the shot-noise limit, or to relax the requirements of the measurement, such as the minimum required detection efficiency. Shaping the statistics of light and using nonclassical optical signals as measurement probes allow for increasing the precision of length measurements \cite{LIGO2013,Pryde2017}, imaging and particle tracking \cite{Brida2010,Bowen2013,Ono2013Sep,Israel2017Mar,Padgett2019}, and spectrophotometry \cite{Krivitsky2016,Matthews2017spectroscopy}. Optical transmittance measurement assisted by correlated photons and single-photon detectors (SPDs) can serve as a prominent example of a quantum-enhanced measurement scheme \cite{Rarity1986,Matthews2017,Genovese2018,Matthews2019}.

The measurement precision at the single-photon level is severely affected by the nonlinearity of the employed photonic detectors. The reason is that the other systematic errors need to be eliminated to reach the quantum regime, while the SPDs themselves maintain strong inherent nonlinearity. One can observe a complex interplay of detector-specific phenomena, such as dark counts, dead time, recovery transition, multi-photon response, and latching. These effects cause highly nontrivial nonlinear behavior that is much stronger compared to classical photodiodes---see Fig.\ref{fig_NL}(B). Not only does the nonlinear response distort the measurement of the average photon flux, it also distorts the measured photon statistics and prevents us from reaching the ultimate precision of quantum-enhanced measurements.
Notable experiments susceptible to detection nonlinearity are the tests of Born's rule in quantum mechanics \cite{Sinha2010,Sollner2012,Magana-Loaiza2016,Kauten2017,Cottere2017,Rengaraj2018}.

Here we explore the nonlinearity of actively and passively quenched single-photon avalanche diodes (SPADs) and a superconducting nanowire single-photon detector (SNSPD) for various bias currents.
The employed nonlinearity measurement does not require a calibrated reference or time-resolved detection, which considerably simplifies the task. 
The nonlinearity characterization is performed with unprecedented accuracy; we reliably detect nonlinearities smaller than 1:1000, and cover seven orders of magnitude of incident illumination.
We discover a supralinear region of SPAD operation, which is not consistent with any known theoretical models and has not been reported yet.
We also characterize the nonlinear behavior of an SNSPD with a complex structure of sub- and supra-linear operation regions. 
We observed, for the first time, supralinearity of a SNSPD under continuous illumination at very low detection rates.

\subsection{Single-photon detectors}
\label{subsection_SPD}

Before discussing the main results, let us briefly review the fundamentals of SPDs and basic principles of nonlinearity characterization.

SPDs convert single-photon absorptions into macroscopic voltages or currents. The current state-of-the-art technology and techniques that are available to achieve single-photon detection are built on various device structures, materials, and non-trivial physical phenomena.
The ones enjoying the most widespread practical use are based on avalanche diodes and superconducting circuits \cite{Natarajan2012,Migdall2013,Chunnilall2014,Hadfield2016,EsmaeilZadeh2021May}. Among these, we investigate detectors that operate in a counting regime (Geiger mode); that is SPADs and SNSPDs.
There are also emerging SPDs based on low-dimensional materials \cite{Rogalski2019, Wang2021}. Such detectors do not yet have established detection models that could consistently model their nonlinearity, which underlines the need to characterize their response by direct measurement.

SPD in a counting regime outputs an electronic pulse when one or more photons are detected \cite{Natarajan2012,Migdall2013,Chunnilall2014,Hadfield2016,EsmaeilZadeh2021May}.
The detection events (counts) arrive at random times with a statistical distribution given by the detected state and the response of the detector. 
The detection rate $R^{\text{det}}$ is then a function of the incident rate $R$; the rate is given in units of counts/s, or simply cps.
Sometimes the detector outputs a pulse even when no photon is detected due to various background contributions (dark counts) or as a result of a previous detection (SPAD afterpulses).
Furthermore, the detector occasionally fails to detect photons because it is not ready to do so after the previous detection event, such as during dead time or a latched state.
Figure \ref{fig_NL}(B) illustrates the nonlinear behavior of a SPD. A simple expectation is a $\int$-shaped sub-linear dependence, where background count rate and dead-time saturation dominate on opposite sides of the power range. However, we found that some detectors exhibit an S-shaped response where the slope gets supralinear in the middle. Supralinearity has been reported for silicon photodiodes for strong classical illumination \cite{Schaefer1983,Tanabe2015,Tanabe2016}, but current SPAD models and measurements do not predict any such phenomenon. 
For SNSPDs, supralinear behavior has been observed for very high rates ($\ge10$~Mcps) due to AC detector coupling \cite{Kerman2013} and for short optical pulses (mean photon number $\ge$ 0.1) as a result of two-photon absorption \cite{Nam2016}. However, no observation of supralinearity of SNSPDs has been reported under continuous illumination with the detection rate below 1~Mcps.

The particular detection imperfections are specific to SPADs and their quenching circuits \cite{Migdall2013,Chunnilall2014}, or SNSPDs \cite{Natarajan2012,Hadfield2016,EsmaeilZadeh2021May} due to their different operational principles.
There is a great number of results in modelling the response of a SPAD with the goal of including all the relevant factors \cite{Stipcevic2013,Kornilov2014,Wang2016,Wayne2017,Straka2020}. Their accuracy has been limited so far and many counter-examples exist for which the measured SPAD response differs from the theoretical model. Consequently, determining the nonlinearity of the SPAD response and finding the optimum detection rate to access the minimum achievable deviation from the ideal linear behavior represents a significant challenge.
This issue is even more pronounced for SNSPDs due to the lack of a precise theoretical model taking into account all physical processes \cite{Nam2016}. A semi-empirical model was proposed and tested with the accuracy $10^{-2}.$\cite{Akhlaghi2009,Lundeen2011}
Detector tomography based on probing with precisely calibrated signals was suggested to thoroughly characterize a detector and obtain the corresponding positive-operator-valued measure \cite{Luis1999,Fiurasek2001,Walmsley2009,Mateo2012,Lundeen2011,WalmsleyHadfield2013,Smith2014,Mateo2017,Andersen2020}. However, if the tomography does not include memory effects, the results can be compromised \cite{Pernice2019}.
The approach presented in the rest of the paper does not rely on a theoretical model or detector tomography. Instead, we focus on a direct measurement of the detector nonlinearity as a function of the detection rate.

\subsection{Nonlinearity characterization}
\label{subsection_NL}

The nonlinearity of various photodetectors, mainly photodiodes, have been explored in great depth using relative and absolute measurement methods \cite{Sanders1972}.
Relative methods require a calibrated reference detector, calibrated attenuators, or time-resolved probe signals and detection.
Absolute measurements, which are generally preferred, are based on a superposition method where the response of the detector is evaluated separately for two signals and their total \cite{Preston1934,Coslovi1980,Schaefer1983,Saunders1984,Haapalinna1999,Shin2013,Weihs2014}.
The individual signals have to be incoherent to prevent optical interference. Often two independent optical sources are used for this reason \cite{Weihs2014}, preferably exhibiting short coherence lengths \cite{Shin2013}. Optionally, the two signals are derived from the same source and superimposed in a slightly misaligned interferometer \cite{Saunders1984,Haapalinna1999}.
Direct absolute measurement of nonlinearity of a SPAD under continuous illumination was reported by Kauten et al., and no statistically significant deviation from the standard model (Eq.~(\ref{SPAD_model}) below) was found \cite{Weihs2014}. We use a single-source approach to directly characterize major single-photon technologies, namely passively and actively quenched SPADs, and SNSPDs for various values of bias current. Our measurements reveal deviations from standard models, as well as anomalous supralinearity.

\section{Direct nonlinearity measurement}
\begin{figure}
	\centering
	\includegraphics[width=\linewidth]{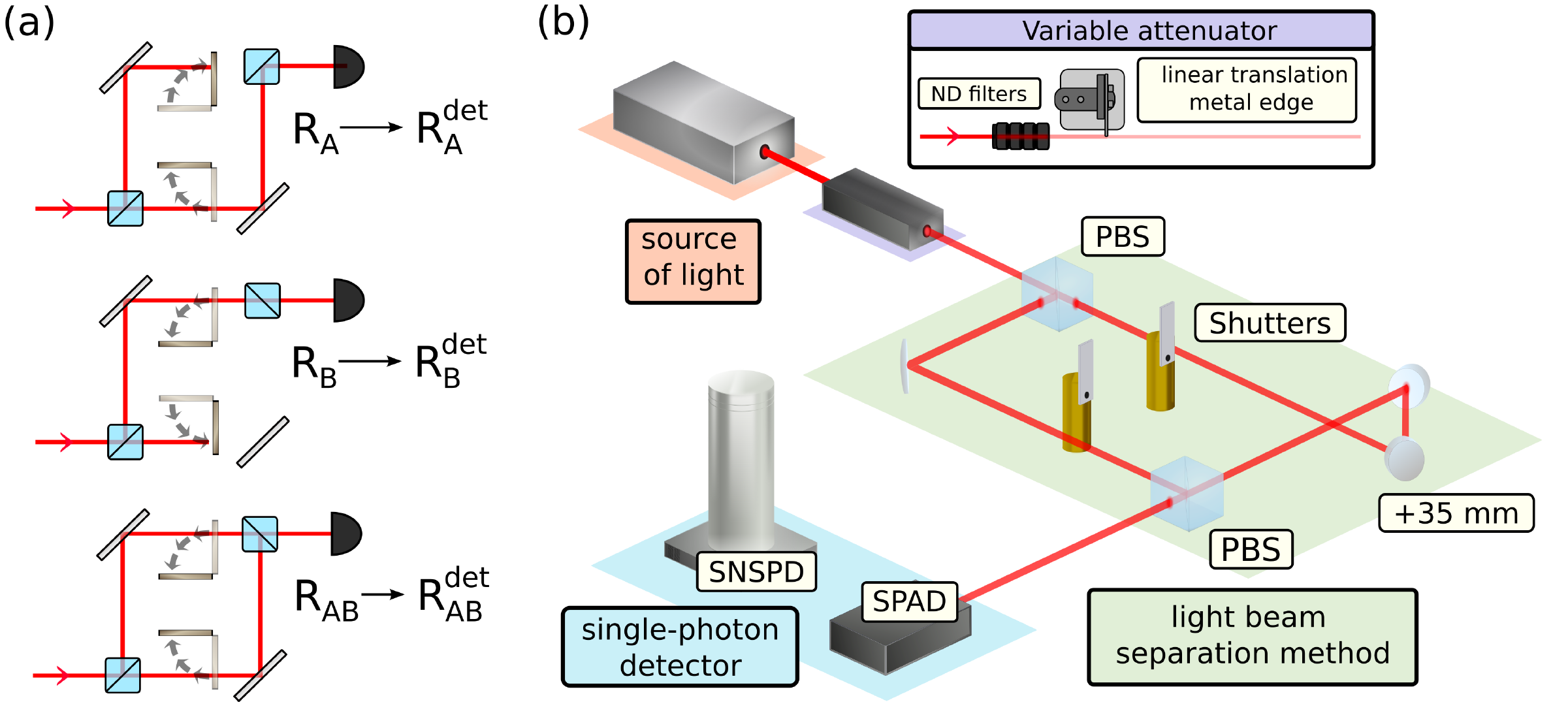}
	\caption{The single-source two-beam method for absolute measurement of nonlinearity consisting of three measurements of the detection rate (A) and the simplified experimental scheme of the method (B).}
	\label{fig_setup}
\end{figure}

We employ the absolute measurement strategy based on the single-source two-beam superposition method.
For a constant intensity level of the optical source, we perform a series of three measurements of the detector response, see Fig.~\ref{fig_setup}(A). The beam is split into two paths A and B that can be individually blocked. The detection rates $R^{\text{det}}_{\text{A}}$ and $R^{\text{det}}_{\text{B}}$ are recorded for each path, and $R^{\text{det}}_{\text{AB}}$ is acquired with both paths open.
The nonlinearity is defined as a deviation from the ideal linear response,
\begin{equation}\label{NL}
	\Delta = \frac{R^{\text{det}}_{\text{A}}+R^{\text{det}}_{\text{B}}}{R^{\text{det}}_{\text{AB}}}-1.
\end{equation}
Our study considers two types of nonlinearity: sub- ($\Delta>0$) and supralinearity ($\Delta<0$). The splitting ratio $R^{\text{det}}_{\text{A}}$\,:\,$R^{\text{det}}_{\text{B}}$ is chosen to be 50:50. For the ultimate nonlinearity characterization, all splitting ratios would need to be measured, but without calibration, the 50:50 splitting is the only one that can be set with certainty, as $R^{\text{det}}_{\text{A}}=R^{\text{det}}_{\text{B}}$.
Furthermore, it harnesses the method's invariance---a 10\% deviation from the 50\% splitting leads to a $\leq 4\%$ relative error in $\Delta$. This is in contrast to using a 3dB attenuator, where a 1\% error in its transmission can change $\Delta$ by an order of magnitude (see the discussion in the Appendix~\ref{sec:attenuator}).

Our approach to the nonlinearity measurement does not require an absolute calibration of a light source.
The measured nonlinearity (\ref{NL}) is a function of the detected count rate (in counts per second). 
This is well-justified, as nonlinearity affects relative measurements of photon flux, where two detection rates are compared, and the corresponding flux ratio needs to be accurately inferred.
In order to correct for the measured value of nonlinearity, we only need to know at which detection rate it occurs.

The experimental setup is shown in Fig.~\ref{fig_setup}(B). A stabilized super-luminescence diode is the source, with the central wavelength 810~nm and the spectral width exceeding  20~nm. The signal is attenuated to scan the input power over the full dynamic range of the detector. To avoid phase interference, the beam is split and joined using polarization beam splitters (PBS), with a 35-mm path difference that is several orders of magnitude larger than the coherence length. The output of the Mach-Zehnder is coupled into a single-mode optical fiber connected to the active area of the SPD under test. The fiber coupling guarantees the same spatial profile of the signals at the detector, but it can decrease the overall stability. Therefore, extra effort was made to reach $10^{-5}$ power stability of the setup including the coupling stages (see Appendix~\ref{sec:stability}).

In the case of SNSPD measurements, the source spectrum was filtered to 12~nm around 800~nm, for which the detector is optimized. The reduced spectral width corresponds to a coherence length shorter than 30~\textmu{}m. The coupling fiber was equipped with a diagonal polarizer and a polarization controller to set the optimal polarization for maximization of the SNSPD efficiency. In this case, the incoherence relied solely on the Mach-Zehnder path difference.

The acquisition time of each individual rate measurement was set to 20~s. The complete characterization of a single detector typically took 20 hours with each measurement repeated 30 times for 40 different rate levels. The precision was found to be limited mostly by fundamental Poissonian variance (see Appendix~\ref{sec:uncertainty}).

We measured three actively quenched SPADs (SPAD-1--3), a passively quenched SPAD (SPAD-P), and an SNSPD for various values of bias current. The output of the detector under test is processed by an FPGA-based counter with a 2.2~ns pulse-pair resolution that is well below the dead time of any detector. For technical information, see Appendix~\ref{sec:setup}.

\section{Results and discussion}

\subsection{Single-photon avalanche diodes}

\begin{figure}[t]
	\centering
	\includegraphics[width=\linewidth]{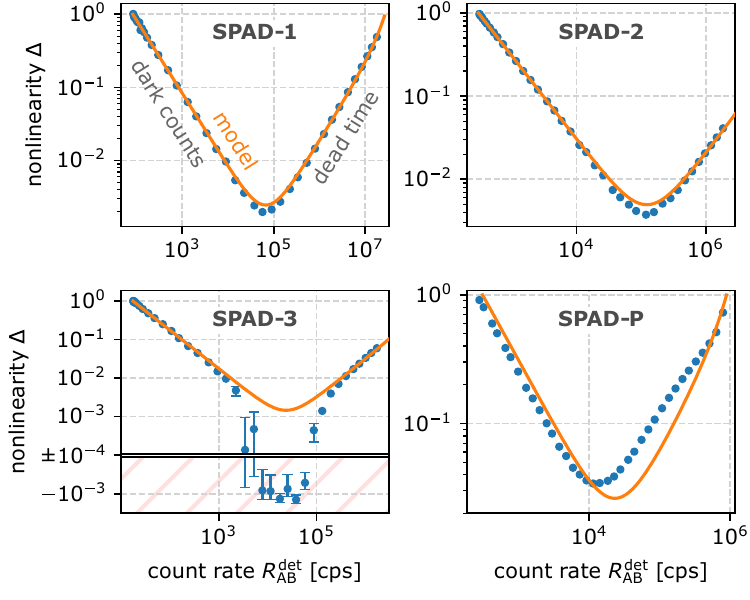}
	\caption{Nonlinear response of the tested SPADs. Solid line represents the theoretical model \eqref{NLmodel}. Each point was measured 30 times and the error bars show the corresponding standard error of the mean.}
	\label{results_SPAD}
\end{figure}

Figure \ref{results_SPAD} shows the measured nonlinearity as a function of the detection rate $R^{\text{det}}_{\text{AB}}$ for various SPADs.
The typical log-log plot of the SPAD nonlinearity $\Delta$ is V-shaped due to dark counts (left slope) and dead time saturation (right slope). The nonlinearity reaches its minimum value ($10^{-3}$ -- $10^{-2}$) for detection rates between $10^4$ and $10^5$ cps.

The established model of the detection rate $R^{\text{det}}$ takes into account the dark count rate $R_0$ and the non-paralyzable dead time $\tau$,\cite{Weihs2014,Straka2020} reading
\begin{equation}
	R^{\text{det}} = f(R) = \frac{R + R_0}{1 + (R+R_0) \tau},
	\label{SPAD_model}
\end{equation}
where $R$ is the incident rate.
The nonlinearity is measured in a balanced configuration, so that the incident rates are $R_\text{A} = R_\text{B} = R_{\text{AB}}/2$. If we substitute the detector model $R^{\text{det}}=f(R)$ given in Eq.~(\ref{SPAD_model}), we obtain the detection rates $R^{\text{det}}_{\text{A}} = R^{\text{det}}_{\text{B}} = f(f^{-1}(R^{\text{det}}_{\text{AB}})/2)$. The expected nonlinearity as a function of detection rate is
\begin{equation}
	\Delta \left(R^\text{det}_{\text{AB}}\right) = \frac{2f\left(\frac{1}{2} f^{-1} \left (R^{\text{det}}_{\text{AB}} \right) \right)}{R^{\text{det}}_{\text{AB}}}-1.
	\label{NLmodel}
\end{equation}
The parameters of this model are the dark count rate $R_0$ and dead time $\tau$.
More elaborate models of actively quenched SPADs are discussed in Appendix~\ref{sec:models}, although none provide a better fit than \eqref{SPAD_model}.

\begin{table}[t]
	\setlength{\tabcolsep}{5pt}
	\centering
	\begin{tabular}{ccccc}
		\hline
		& SPAD-1 & SPAD-2 & SPAD-3 & SPAD-P  \\	\hline
		$R_{0}$ [cps]  & 88(3) & 314(6) & 20(2) & 264(1)  \\
		$R_{0,\text{fit}}$ [cps] & 83(4) & 304(2) & 17.5(7) & 300(200)  \\
		$\tau_{\text{R}}$ [ns] &  29.5(5)    &   47.0(5)   &  56.6(4)    &   517(6) \\
		$\tau_{\text{fit}}$ [ns] &   36.7(1)   &    40.2(4)  &  61(1)    &   1130(20) \\
		\hline
	\end{tabular}
	\caption{Comparison of directly measured dark counts $R_{0}$ and  recovery times $\tau_{\text{R}}$ that were measured directly using time-resolved detection techniques \cite{Straka2020}, and the dead time values $\tau_{\text{fit}}$ and dark counts $R_{0,\text{fit}}$ that were the best fit of the model (\ref{SPAD_model}).}
	\label{table}
\end{table}

The data were fit with Eq.~(\ref{NLmodel}) in Fig.~\ref{results_SPAD}, and they are in significant disagreement with the model, affirming the need for direct nonlinearity measurement. The most prominent feature is the supralinear behavior ($\Delta < 0$) of SPAD-3 in Fig.~\ref{results_SPAD}, which is a hitherto unreported phenomenon for SPADs. In fact, all actively quenched SPADs 1--3 exhibit systematically lower nonlinearities than expected in their minima, and supralinearity appears after correcting for $R_0$ and $\tau$ (see Fig.~\ref{fig_datacorrected} in Appendix~\ref{sec:models}). Additionally, the fit parameters $R_0, \tau$ do not agree with values that were obtained from independent time-resolved measurements.

A comparison of dead times and directly observed recovery times are given in Table~\ref{table}. For actively quenched detectors (SPAD-1--3), the measured recovery time $\tau_{\text{R}}$ also includes a brief detector reset time, and so must always be $\tau_{\text{R}} \geq \tau$.\cite{Ware2007} Reset effects like twilight pulsing could therefore explain the difference for SPAD-2, but not for the other detectors SPAD-1 and SPAD-3 \cite{Straka2020}. SPAD-P is passively quenched and so the recovery and dead time values differ significantly. Such detectors exhibit gradual efficiency recovery after each detection, which would require a complex empirical model. As shown in Appendix~\ref{sec:models}, no other few-parameter models offer a better fit than (\ref{SPAD_model}).

\begin{figure}
	\centering
	\includegraphics[width=\linewidth]{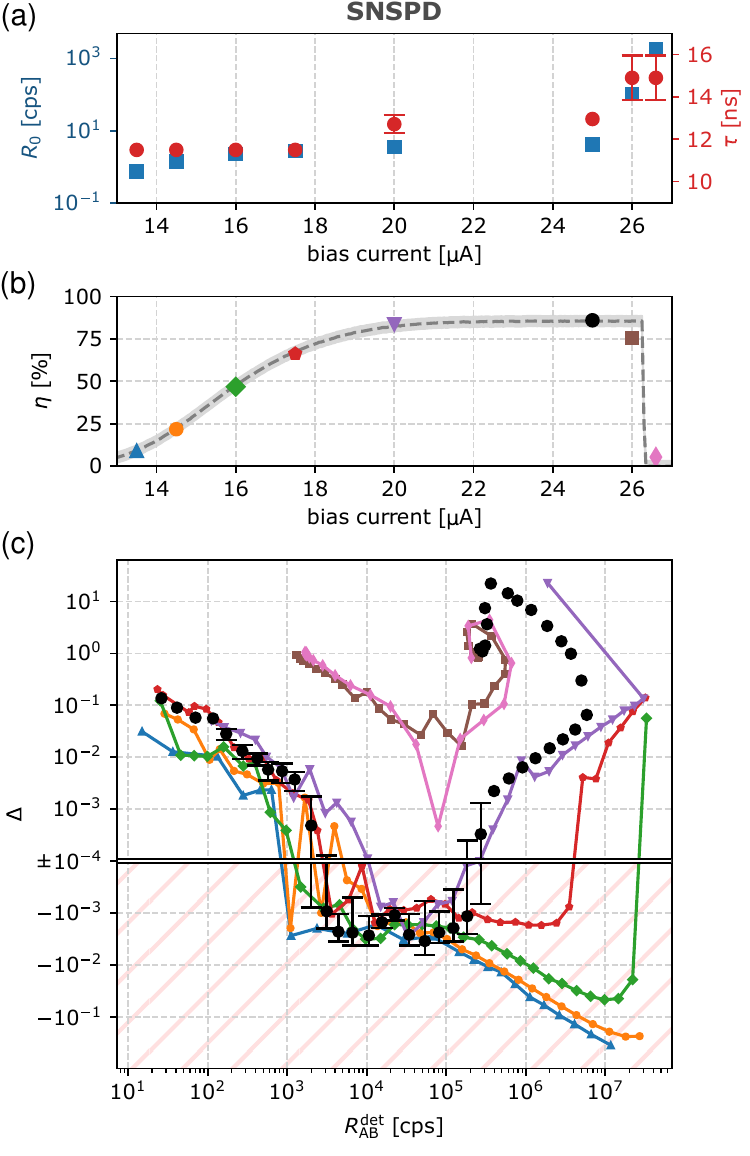}
	\caption{(A) Dark counts $R_0$, dead time $\tau$ and (B) detection efficiency $\eta$ as a function of bias current. Dashed line is the manufacturer's specification. (C) the measured nonlinearity of the SNSPD for the values of bias current given by the respective markers in (B).}
	\label{results_SNSPD}
\end{figure}

Because not all systematic errors can be seen in Fig.~\ref{results_SPAD}, we evaluated the $\chi^2$ with the highest $p$-value being $p_\text{SPAD-2} \sim 10^{-70}$. It means that under the assumption that the model (\ref{SPAD_model}) is valid and only the statistical error takes effect, the probability of obtaining the measured data or any worse data would be equal to the $p$-value. This means that the model (\ref{SPAD_model}) already deviates from the measurement with high statistical significance for 20~s integration times. Additionally, we applied standard rate corrections on the data (inversion of \eqref{SPAD_model}) to evaluate any residual nonlinearity, which reveals significant mismatch even for $R_{\text{AB}}^{\text{det}} \gtrsim 10^6$~cps. A detailed statistical analysis can be found in Appendix~\ref{sec:models}.

We have also tested more elaborate response models such as Ref.~\onlinecite{Straka2020} (including dead time, afterpulses, and twilight pulses), and a combination of paralyzable and non-paralyzable dead time \cite{Feller1948,Evans1955,Knoll1989,Muller1991,Gardner1997,Lee2000} developed originally for Geiger--M\"uller detectors. None of these can reproduce the measured results or explain supralinearity, which is shown in Appendix~\ref{sec:models}. We also ruled out potential changes in dead time, dark counts, efficiency, or afterpulsing as hypothetical explanations for the observed supralinearity (see Appendix~\ref{sec:models}).

Current evidence shows that SPADs do not always exhibit constant dark counts \cite{Karami2010}, efficiency \cite{Raupach2022Apr}, or recovery time (dead time + reset time) \cite{Straka2020}. This means that state-of-the-art models are insufficient, leaving the experimentalist with two options: 1. a detailed characterization of the SPAD involving time-resolved measurements and circuit analysis; 2. direct nonlinearity measurement.

\subsection{Superconducting nanowire single-photon detectors}

Figure \ref{results_SNSPD} shows nonlinearity of the SNSPD with respect to the bias current. Subfig.~\ref{results_SNSPD}(A) shows the dependence of dead time and dark counts on the bias. The property affected the most is detection efficiency, shown in subfig.~\ref{results_SNSPD}(B). The detection efficiencies were calculated relatively to the manufacturer's specification of $86\pm 3 \%$ for 25~\textmu{}A.
Subfig.~\ref{results_SNSPD}(C) shows the measured nonlinearity $\Delta$ as a function of the total detected rate $R^{\text{det}}_{\text{AB}}$. Each data set corresponds to a different bias current, where the plot markers match the respective points in subfig.~\ref{results_SNSPD}(B).

The nonlinearity of the SNSPD is a combination of several phenomena. For lower rates, the effect of dark counts is easily recognizable.  Around $10^4$~cps, all sub-critical regimes ($I_{\text{bias}} \leq 25$~\textmu{}A) begin to exhibit supralinearity ($\Delta < 0$). For higher count rates, two scenarios are observed. Bias currents corresponding to the efficiency plateau (20--25~\textmu{}A) lead to dead time saturation, while lower biases maintain the supralinearity. Finally, for $R^{\text{det}}_{\text{AB}} > 10^6$~cps, latching rapidly increases for all but the lowest biases. This introduces strong saturation and eventually results in an inverse detection response, where count rate decreases with increasing illumination.

Let us address the supralinear behavior, linked previously to two effects. With decreasing bias current, two-photon and higher-order detection efficiencies rise as indicated by detector tomography \cite{Lundeen2011}. So far, the two-photon absorption of SNSPDs was observed for ultra-short optical pulses in picosecond regime with mean photon numbers $\ge$ 0.1.\cite{Nam2016}
Under continuous illumination, the multi-photon absorption becomes challenging to model due to hotspot relaxation dynamics \cite{Nam2016}. The second factor is the AC coupling of the readout circuit---the settling of the bias current after each detection results in rate-dependent efficiency. The supralinear behavior due to the AC coupling was observed only for very high rates $\ge10$~Mcps.\cite{Kerman2013}
Our results represent the first observation of the SNSPDs supralinearity under continuous illumination with the detection rates as low as $10^3$~cps.

\section{Conclusion and outlook}

We designed and realized a direct measurement of single-photon detection absolute nonlinearity with high accuracy over several orders of magnitude of light intensity. We performed the measurement for SPADs and SNSPDs. The measurement technique does not require a calibrated single-photon source, a calibrated reference detector, or time-resolved detection. Due to the method's robustness, the nonlinearity is not affected by technical issues, but rather reflects the detector properties.

For all SPADs, we found significant disagreement of established theoretical models with the measured data. We also discovered anomalous supralinear behavior of a SPAD. This phenomenon has been neither predicted nor observed yet, and cannot be explained in terms of the known SPAD parameters. The SPADs in question were based on silicon, but SPADs based on III--V materials (InGaAs/InP) possess the same principle of operation and therefore exhibit similar nonlinear phenomena as silicon SPADs, albeit of different magnitudes. We expect that sublinear behavior would be generally stronger for the InGaAs/InP SPADs than silicon SPADs due to higher values of dark counts and dead time. Consequently, the chance of observing subtle deviations such as supralinearity is expected to be lower.

For the SNSPD, we performed its detailed nonlinearity analysis over 7 orders of magnitude of incident illumination and the full range of relevant values of bias current, which goes beyond any SNSPD characterization reported so far.
We detected SNSPD supralinearity under continuous illumination at unprecedentedly low detection rates down to $10^3$ cps, which has not been observed before.
The results for both SPADs and SNSPDs show that nonlinearity in single-photon detection is a complex mixture of non-trivial phenomena which have so far eluded accurate theoretical description.

Direct SPD characterization can also be applied to emerging detection techniques, such as those based on low-dimensional materials. The results in this field suggest susceptibility to nonliner response as well.

Some of these detectors operate in a linear gain regime, such as p-n junction arrays \cite{Gibson2019May}, 2D-layer Schottky barriers \cite{Lei2015}, or quantum dots \cite{Bulgarini2012,Gansen2007}. The results in Refs.~\onlinecite{Gibson2019May,Lei2015,Bulgarini2012} show that the overall nonlinearity can be quite significant and dependent on the bias voltage. Until there are direct nonlinearity characterizations performed for such detectors, it is hard to speculate whether both sub- and supra-linearity could be expected, but the published results suggest that both can be present.

Low-dimensional detectors based on photogating operate in a counting regime \cite{Luo2018,Roy2017}. That means individual single-photon absorptions are distinguished as quantized jumps in photocurrent. Due to the long-lived traps responsible for the photogating effect, the trapped carriers can accumulate even at low rates below 1~cps. Additionally, trap decays cause reverse jumps in photocurrent that can, in principle, interfere with detection readouts \cite{Luo2018}. Electronic noise contributes to cross-talk in photon-number distinction. All these effects may contribute to nonlinearity, and the results in Ref.~\onlinecite{Luo2018} suggest both strong sub- and supra-linearity.

A Josephson-junction-based detector reported in Ref.~\onlinecite{Walsh2021} also operates in a counting regime. The results therein only show nonlinearity due to dark counts, but a dedicated measurement may reveal additional effects.

There are other photosensitive nanostructures\cite{Gao2019} that have not yet been sufficiently characterized in terms of their response to incident power. With all such emerging single-photon detection techniques, the ability to perform direct self-contained characterizations is valuable due to the lack of accurate theoretical or empirical models.

The findings presented in this work can be applied to radiometric, spectroscopic, imaging, and optical communication methods that rely on precise assessment of illumination or transmission levels. We have shown that correcting for nonlinear aspects based on state-of-the-art response models is generally not sufficient, and so direct measurement of nonlinearity becomes a necessity.
Accurate detector calibration is particularly critical for measuring photon statistics and reaching the quantum advantage in quantum metrology.
The presented results open the way for ultra-precise identification and mitigation of nonlinear effects in SPDs.

\section*{Acknowledgments}
We acknowledge the funding from the Czech Science Foundation (project 21-18545S),
MEYS and European Union's Horizon 2020 (2014--2020) research and innovation framework programme (project HYPER-U-P-S, No.~8C18002).
Project HYPER-U-P-S has received funding from the QuantERA ERA-NET Cofund in Quantum Technologies implemented within the European Union's Horizon 2020 Programme.
J.H. acknowledges the Palack\'y University projects IGA-PrF-2021-006 and IGA-PrF-2022-005.

\section*{Conflict of interest statement}
The custom-made counter used in the presented measurements is commercially available and sold by the Palack\'y University. M.J. is one of the developers and thus eligible for a percentage of the income. However, the presented measurements are not affected by the choice of the counter, as the necessary parameters can be met by other devices available on the market. All other authors declare they have no competing interests.

\section*{Data availability statement}
The data that support the findings of this study are openly available on GitHub, reference number \onlinecite{HlousekGitHub2021}.

\appendix

\section{Experimental setup}
\label{sec:setup}

\begin{figure*}[ht]
	\centering
	\includegraphics[width=1.0\textwidth]{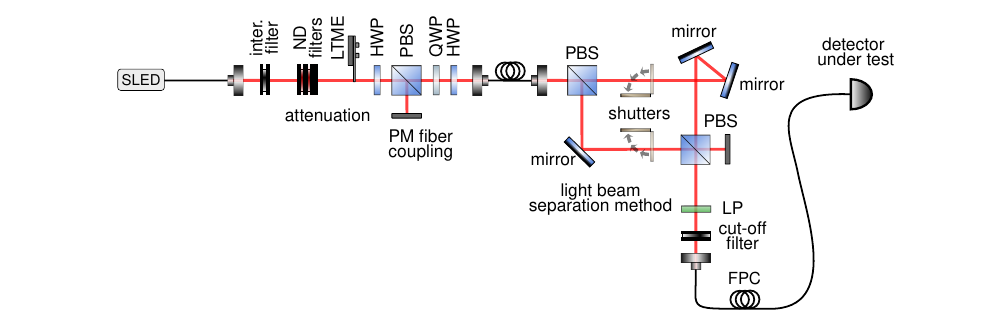}
	\caption{
		Experimental setup of the single-source two-beam superposition method for absolute nonlinearity measurement: the scheme includes preparation of a stabilized and attenuated optical signal, its polarization control, beam separation and switching, and detector under test.
	}
	\label{fig_S1}
\end{figure*}

Figure \ref{fig_S1} depicts the experimental setup for absolute measurement of nonlinearity using the single-source two-beam superposition method. The light source is a temperature-stabilized super-luminescence diode (QPhotonics QSDM-810-2) in constant current mode with the central wavelength of 810 nm and the spectral width exceeding 20~nm.
Angled physical contacts (APC) between optical fibers reduce back-reflections approximately by 60~dB and improve the stability of the source. Attenuation by several orders of magnitude is needed to generate an optical signal in the dynamic range of the tested detector. For this purpose, the optical beam is strongly attenuated by a series of neutral density (ND) filters (Thorlabs NE530, NE520, NE513, and NE504) down to $\sim 10^7$ counts per second. Continuous adjustment is realized by a metal edge fixed to a linear translation motorized stage (LTME, Newport MFA-CC) driven by a motion controller (SMC-100CC). The moving edge along with single-mode fiber coupling serves as variable attenuation.

\begin{figure*}[ht]
	\centering
	\includegraphics[width=0.9\textwidth]{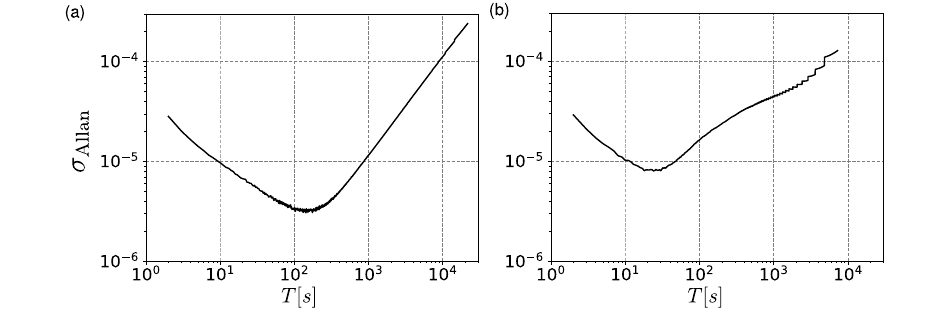}
	\caption{
		Optical intensity fluctuations represented by the relative Allan deviation. 
		Shown are: (a) temperature-stabilized super-luminescence diode (QPhotonics QSDM-810-2) in constant current mode, and (b) signal from the same source transmitted through the experimental setup for nonlinearity measurement.
	}
	\label{fig_S2}
\end{figure*}

Since the splitting ratio of the interferometer is sensitive to input polarization, the input beam is well polarized and coupled in the slow axis of a polarization-maintaining fiber.
The unbalanced Mach-Zehnder interferometer consists of two polarizing beam splitters (PBS) with extinction ratio $T_{\text{P}}:T_{\text{S}} > 1000:1$. The Mach-Zehnder setup is used to split the attenuated beam into two spatially separated beams that can be individually blocked and superimposed again at the output.
The combination of perpendicular polarizations and the path difference $l=35.4$~mm between the individual arms ensure incoherent mixing at the second polarization beam splitter. 
Mechanical blocking of the two optical signals is done by compact home-made optical shutters using a digital RC servo (Savox SH-0262MG).
Thin (0.5~mm) metal flag with dimensions 15~$\times$~70~mm attached to the servo shaft blocks the collimated Gaussian beam that has a radius $w_{\text{0}} = 1.03$~mm.
To perform the operation of opening or blocking the beam, it is necessary to turn the servo by $15^{\circ}$ with a total latency of 50~ms.
The RC servo is controlled by a pulse-width modulation signal generated by Arduino Uno with a microprocessor AT-mega328.
For more information about the employed shutters, see Ref.~\onlinecite{QOLOshutters}.
After the second PBS, the output light is fed into a single-mode fiber and coupled to the tested detector.
Stray light is eliminated by a cut-off filter (Semrock BLP01-635R-25).

We investigated the response of five SPDs: actively (SPAD-1--3) and passively quenched (SPAD-P) thick-junction silicon single-photon avalanche diodes (SPADs), and a NbTiN superconducting (2.7~K) nanowire single-photon detector (SNSPD) with AC readout and room-temperature amplifier. Namely,
SPAD-1 -- Excelitas SPCM-AQRH CD3432H,
SPAD-2 -- Perkin Elmer SPCM-AQ4C module s.n.\@ 167,
SPAD-3 -- Laser Components COUNT-20C-FC D4967,
SPAD-P -- ID Quantique ID120 s.n.\@ 1518006, and
SNSPD -- Single Quantum Eos CS SNSPD system s.n.\@ SQ071.
The detectors differ in many technical parameters and physical effects such as active area, photon detection efficiency, dark count rate, dead time, and afterpulsing probability. The manufacturer’s specifications are shown in Table \ref{table_1}.
The detector SPAD-P (ID Quantique ID120) differs from SPAD-1--3 not only by the passive quenching mechanism, but also by being a free-space module with tunable temperature and bias voltage.
All measurements were done at the temperature $-40\,{}^{\circ}$C and bias voltage 180~V.

The SNSPD possesses a limited spectral region of the maximum efficiency. 
For this reason, the wide emission of SLED was reduced using a 12~nm band-pass interference filter (Semrock FF01-800/12-25).
Furthermore, the SNSPD is naturally sensitive to polarization due to its nanowire structure.
The optical beams of the unbalanced Mach-Zehnder interferometer have orthogonal H/V polarizations, so it is necessary to add a diagonally oriented linear polarizer (LP) at the output to make them indistinguishable for the detector.
A fiber polarization controller (FPC) is used to optimize the detection efficiency with respect to polarization.

Electronic output signals from the tested detectors were processed by an electronic counter.
The most critical parameter is its pulse-pair resolution, which has to be better than the recovery time of all tested detectors. Initially, we were using a commercial 100~MHz counter (ORTEC 974C), which was used to measure the SPADs.
Later, we developed an FPGA-based 230~MHz counter with a 2.2~ns pulse-pair resolution and digitally tunable threshold voltages (QOLO Countex \cite{QOLOcountex}). This counter was used to perform measurements on the SNSPD, as well as additional measurements on SPAD-1 and SPAD-3. Detectors SPAD-1 and SPAD-3 were measured using both counters with the same results.

\begin{table}[ht]
	\centering
	\caption{
		Manufacturer specifications for tested detectors. Parameters shown: active area diameter $A$, photon detection efficiency $\eta$, dark count rate $R_{0}$, dead time $\tau$, and afterpulsing probability $p_{\text{a}}$.
	}
	\label{table_1}
	\begin{ruledtabular}
		\begin{tabular}{cccccc}
			& $A$ [\textmu{}m] & $\eta$ [\%] & $R_{0}$ [cps]& $\tau$ [ns]& $p_{\text{a}}$ [\%] \\
			\hline
			SPAD-1 & 180 & 50 & $<$100 & 28.7 & 0.1  \\
			SPAD-2 & 180 & 47 & $<$500 & 50 & 0.3  \\
			SPAD-3 & 100 & 57 & $<$20 & 55 & 0.31  \\
			SPAD-P & 500 & 80 & 200 & 1000 & negligible \\
			SNSPD & -- & 86 & $\leq$10 & $\leq$10 & none \\
		\end{tabular}
	\end{ruledtabular}
\end{table}

\section{Stability of the source and measurement setup}
\label{sec:stability}

Generally, intensity fluctuations of the light source can affect the measurement accuracy.
We investigated the stability of the SLED and also the intensity stability at the output of the setup.
We performed a long-term measurement of optical intensity and evaluated the Allan deviation \cite{Allan1966} to determine the integration time $T$ for which the measurement is least affected by intensity fluctuations.
Figure \ref{fig_S2} shows the relative Allan deviation $\sigma_{\text{Allan}}$ as a function of the integration time $T$.
The deviation is better than $10^{-5}$ for integration times from $10$~s to $10^{3}$~s. 
For the nonlinearity measurement setup, optimal integration times are shifted toward shorter times and the minimum relative Allan deviation gets worse, but still below $10^{-4}$ for integration times up to $6\times10^{3}$~s.
The raw data and their processing are available on GitHub \cite{HlousekGitHub2021}.

\section{Measurement uncertainty}
\label{sec:uncertainty}

\subsection{Shot noise limitation}

We performed an analysis of nonlinearity measurement uncertainty $\sigma(\Delta)$ to find its ultimate physical limits and find out whether the data approach the fundamental nonlinearity resolution.
The statistical uncertainty was calculated for the case of a balanced experimental setup $R^{\text{det}}_\mathrm{A}=R^{\text{det}}_\mathrm{B}$, measurement time $T$ = 20 s for each rate, and $N=30$ repeated measurements.

The fundamental limit is represented by the shot noise of the Poisson counting process, where the measured number of events ($R^{\text{det}}T$) exhibits variance equal to the mean value. This implies the standard deviation of the detection rate $\sigma(R^{\text{det}}) = \sqrt{R^{\text{det}}/T}$. If we introduce this standard deviation to each detection rate in the nonlinearity formula
\begin{equation}\label{NL_app}	
	\Delta = \frac{R^{\text{det}}_{\text{A}}+R^{\text{det}}_{\text{B}}}{R^{\text{det}}_{\text{AB}}} - 1,
\end{equation}
through standard error propagation $\sigma^2(\Delta) = \sum_i \left[\sigma(R^{\text{det}}_i) \times \partial \Delta/\partial R^{\text{det}}_i\right]^2$, we obtain the standard deviation
\begin{equation}
	\sigma(\Delta) = \sqrt{\frac{(1+\Delta)(2+\Delta)}{R^{\text{det}}_{\text{AB}}T}} \approx \sqrt{\frac{2}{R^{\text{det}}_{\text{AB}}T}}.
\end{equation}
The measured nonlinearity $\overline{\Delta}$ is averaged from $N$ measurements, so the standard deviation of the result is
\begin{equation}
	\sigma \left(\overline{\Delta}\right) = \sigma(\Delta)/\sqrt{N}.
\end{equation}

\begin{figure*}[ht]
	\centering
	\includegraphics[width=1.0\textwidth]{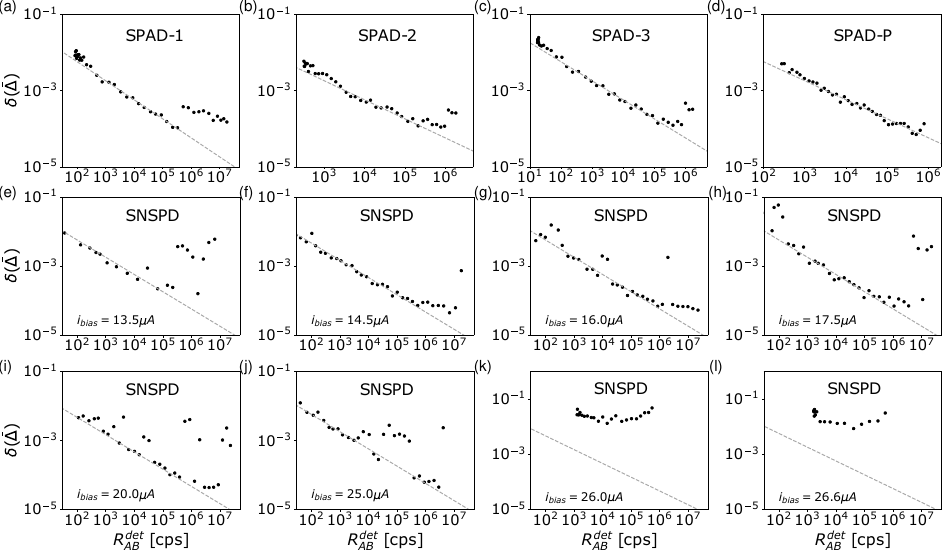}
	\caption{
		The nonlinearity measurement standard error $\sigma(\overline{\Delta})$ of tested detectors (a) SPAD-1, (b) SPAD-2, (c) SPAD-3, (d) SPAD-P and (e-f) SNSPD as a function of detection rate $R^{\text{det}}_{\text{AB}}$.
		Subplots (e-f) show $\sigma(\overline{\Delta})$ of an SNSPD with respect to the bias current.
		The measured standard errors are plotted as black dots, and the lower bound on the standard deviation is represented as a grey dashed line.
	}
	\label{fig_S3}
\end{figure*}

\begin{figure}
	\centering
	\includegraphics[width=\linewidth]{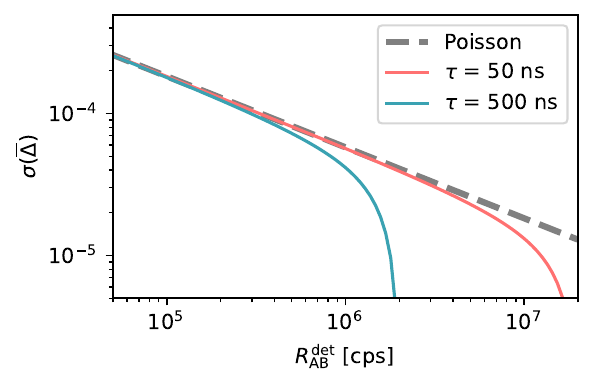}
	\caption{The effect of dead-time saturation on the fundamental limit of uncertainty resolution. The integration time is $T$ = 20 s and the number of measurements $N=30$.}
	\label{fig:uncertaintydeadtime}
\end{figure}

This limit is compared to the experimental values in Fig.~\ref{fig_S3}.
The measured standard errors of the nonlinearity for SPADs meet their lower bound $10^{-4}$ for rates higher than $2\times10^{5}$~cps (Fig.~\ref{fig_S3}(a-d)). For higher rates, excess noise limits the measurement uncertainty; it nevertheless remains below $\sigma(\overline{\Delta}) < 10^{-3}$. In this region, the mean nonlinearity value is larger than the uncertainty by at least one order of magnitude.
Subfigures (e-l) show the nonlinearity measurement uncertainty of the SNSPD for several different values of bias current.
Latching rapidly increases for bias current $I_{\text{bias}} > 25$~\textmu{}A and brings extra uncertainty to the measurement (Fig.~7(k) and (l)).

The shot noise limit is well applicable to Poisson count rates. However,  dead-time saturation will further affect the measurement, because the measured count rates become sub-Poissonian as they approach the limit $R^{\text{det}} \to 1/\tau_{\text{NP}}$. Because the standard deviation of accumulated counts is complicated to calculate analytically \cite{Straka2020}, we are going to assume a sufficiently long integration time so that the number of accumulated counts is large: $\langle n \rangle = R^{\text{det}}T \gg 1$. This is a reasonable assumption for high rates, where saturation occurs. Then, we can make the following derivation.

Upon measuring $n$ detections during a time $T$, we obtain the average time between detections, $\overline{\Delta t}= T/n$. Because each realization of $\Delta t$ is independent, the average time $\overline{\Delta t}$ is normally distributed with a standard deviation $\sigma(\overline{\Delta t}) = \sigma(\Delta t)/\sqrt{n}$ according to the central limit theorem. We assume $\Delta t$ to be a dead time $\tau_{\text{NP}}$ plus an exponentially distributed delay with a rate parameter $\lambda$, which makes $\langle \Delta t \rangle = \tau_{\text{NP}} + 1/\lambda$ and $\sigma(\Delta t) = 1/\lambda$. Because $n \gg 1$, the standard deviation is small relative to the mean; $\sigma(\overline{\Delta t}) \ll \langle \Delta t \rangle$. As a result, the mean rate $R^{\text{det}} = 1/\overline{\Delta t}$ is also normally distributed and the standard deviation is scaled based on the slope around the mean value,
\begin{equation}
	\sigma(R^{\text{det}}) = \left| \frac{\partial R^{\text{det}}}{\partial \overline{\Delta t}} \right| \sigma(\overline{\Delta t}) = (R^{\text{det}})^2 \sigma(\overline{\Delta t}).
\end{equation}
Due to the large number of detections, on the right side we can simplify $\overline{\Delta t} \approx \langle \Delta t \rangle \approx 1/R^{\text{det}}$, so
\begin{equation}\label{eq:rate_uncertainty}
	\sigma(R^{\text{det}}) = (1-\tau_{\text{NP}} R^{\text{det}})\sqrt{R^{\text{det}}/T}.
\end{equation}

The resulting sub-Poissonian bounds are shown in Fig.~\ref{fig:uncertaintydeadtime} for two dead time values to illustrate the cases of an actively and passively quenched detectors. The bounds do not make a difference within the data range of Fig.~\ref{fig_S3}.

\subsection{Optimal measurement times}

Here we address the optimal allocation of measurement time to minimize the uncertainty. Given the overall measurement time $T_O$ per one sample of $\Delta$, there exists an optimal distribution between measuring $R^{\text{det}}_{\text{A}}$, $R^{\text{det}}_{\text{B}}$, and $R^{\text{det}}_{\text{AB}}$. Assuming a balanced splitting $R^{\text{det}}_{\text{A}} \approx R^{\text{det}}_{\text{B}}$, the minimum uncertainty of $\Delta$ is reached for
\begin{align}
	T_{\text{AB}} &= \frac{T_O}{1+\sqrt{2/(1+\Delta)}},	\\
	T_{\text{A}} &= T_{\text{B}} = \frac{1}{2} \left(T_O - T_{\text{AB}} \right).
\end{align}
In the regime where $|\Delta| \ll 1$, the approximate ratios are $T_{\text{A}} : T_{\text{B}} : T_{\text{AB}} = 0.3 : 0.3 : 0.4$. The assumption we made on the way is that the measurement time is long enough so that all the accumulated counts $C_i$ ($i=\text{A, B, AB}$) are approximately normally distributed. This holds if $\langle C_i \rangle = R^{\text{det}}_iT_i \gg 1$.

\section{Two-beam method compared to an attenuator}
\label{sec:attenuator}

Here we illustrate the advantage of the two-beam method compared to single-beam measurements. An analogous way of characterizing nonlinearity is comparing the detection rates of two intensity levels by employing an attenuator of power transmittance $\eta$. The nonlinearity $\Delta$ essentially tests two quantities that should be equal if the detector was linear; $R^{\text{det}}_{\text{A}} + R^{\text{det}}_{\text{B}} \stackrel{?}{=} R^{\text{det}}_{\text{AB}}$. With an attenuator, we can only measure the transmitted power $R^{\text{det}}_\eta$ and total power $R^{\text{det}}_1$, whereas the lost power $R^{\text{det}}_{1-\eta}$ is inaccessible. We can, however, analogously test $R^{\text{det}}_{1-\eta} \stackrel{?}{=} \frac{1-\eta}{\eta} R^{\text{det}}_\eta$. Thus, we construct nonlinearity

\begin{equation}\label{eq:Delta_alt}
	\Delta^{\text{att}} = \frac{R^{\text{det}}_\eta + \frac{1-\eta}{\eta} R^{\text{det}}_\eta}{R^{\text{det}}} - 1 = \frac{1}{\eta}\frac{R^{\text{det}}_\eta}{R^{\text{det}}} - 1.
\end{equation}
For $\eta = 1/2$, this is equal to nonlinearity $\Delta$ \eqref{NL}.

Let us examine a scenario where we attempt to measure nonlinearity for a balanced scheme, but there is a slight error in calibration,
\begin{equation}
	\eta = 0.5 + \delta\eta.
\end{equation}

Because the value of nonlinearity may span many orders of magnitude, let us examine its \emph{relative} error, meaning the difference proportional to the ideal value of $\Delta$ for $\eta = 0.5$:
\begin{equation}
	\delta\Delta = \left| \frac{\Delta(0.5 + \delta\eta) - \Delta(0.5)}{\Delta(0.5)} \right| .
\end{equation}
This will generally depend on the model of the detector and the incident rate, so let us illustrate the basic behavior for a simple non-paralyzing model with background $R_0$ and dead time $\tau$
\begin{equation}
	R^{\text{det}}_\eta := R^{\text{det}}(\eta R) = \frac{1}{\frac{1}{\eta R+R_0}+\tau}.
\end{equation}
The nonlinearity for the two-beam measurement is then
\begin{equation}
	\Delta(\eta) = \frac{R^{\text{det}}_\eta + R^{\text{det}}_{1-\eta}}{R^{\text{det}}_1} - 1, 
\end{equation}
while for the attenuator, an unknown error manifests only in the measured intensity,
\begin{equation}
	\Delta^{\text{att}}(\eta) = \frac{1}{0.5}\frac{R^{\text{det}}_\eta}{R^{\text{det}}_1} - 1.
\end{equation}
Apart from $\eta$, these quantities depend on $R$, $R_0$, and $\tau$. When examining the error, we look for the maximum over all incident rates $R$. Thus, we obtain the maximum relative error (for both nonlinearities)
\begin{equation}
	\delta\Delta_{\text{max}}(\delta\eta) = \max_R \left| \frac{\Delta(0.5+\delta\eta)}{\Delta(0.5)} - 1 \right| .
\end{equation}

Now, let us assume a small error in calibration $\delta\eta \ll 1$, so we can consider polynomial contributions of $\delta\eta$. Let us also examine the quantities for a certain realistic range of parameters,
\begin{align}
	0 < R_0 &< 10^3\text{ cps}, \\
	0 < \tau &< 10^{-6}\text{ s}.
\end{align}

For $\delta\Delta^{\text{att}}_{\text{max}}$, we make certain approximations, chiefly $R_0 \tau \ll 1$, that yield the maximal rate $R \approx \sqrt{2 R_0 / \tau}$ and
\begin{equation}
	\delta\Delta^{\text{att}}_{\text{max}} \approx \sqrt{\frac{2}{R_0 \tau}} \times \delta\eta.
\end{equation}

On the other hand, the two-beam nonlinearity error scales more favorably, and with a constant factor,
\begin{equation}
	\delta\Delta_{\text{max}} \approx 4 \times \delta\eta^2.
\end{equation}

In practical terms, for a 1\% deviation in beam splitter transmittance, $\delta\eta = 0.01$, the value of nonlinearity measured by the two-beam method will change by $\delta\Delta \lesssim 0.04\%$ for any realistic dead time and dark counts.

If the same 1\% error is applied to an attenuator, the error depends on the detector parameters. For an extreme case of $R_0 = 10^3$~cps and $\tau = 1$~\textmu{}s, the error is $\delta\Delta^{\text{att}} \lesssim 44\%$, whereas for values corresponding more with modern SPADs, $R_0 = 50$~cps and $\tau = 30$~ns, the value can change drastically by  $\delta\Delta^{\text{att}} \lesssim 1200\%$.

By accessing the complementary optical power, the two-beam nonlinearity offers a much more reliable measurement. The dependence on the splitting ratio is quite weak, with the value of $\Delta$ changing by $\lesssim 15\%$ within the range $\eta \in (0.3, 0.7)$ relative to $\eta = 0.5$. Furthermore, the balanced splitting ratio is the only one that can be calibrated with arbitrary precision, so it easily becomes the most suitable measurement setting to use, as it represents the nonlinear properties of the detector much more than any technical parameters of the measurement.

\section{Repeatability of the measurement}
\label{sec:repeatability}

\begin{figure}[ht]
	\centering
	\includegraphics[width=\linewidth]{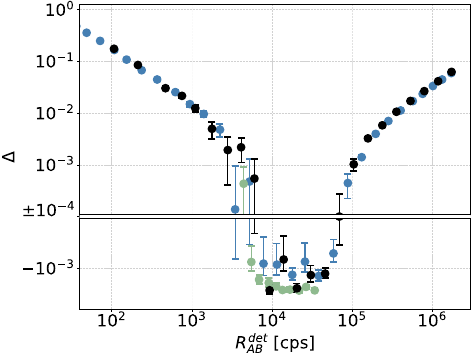}
	\caption{
		The repeatability of the nonlinearity measurement of the detector SPAD-3 throughout the years. 
		The nonlinearity $\Delta$ is a function of the detection rate $R^{\text{det}}_{\text{AB}}$, shown at different dates: black -- February 2021, green -- January 2016, and blue -- November 2014.
	}
	\label{fig_count}
\end{figure}

In 2014 we built the experimental setup for absolute measurement of nonlinearity to characterize the detector response and since then, we have performed nonlinearity measurements repeatedly.
As an example representing the consistency of the results obtained from the presented method, we chose the actively quenched SPAD from Laser Components (SPAD-3). 
This detector exhibits anomalous supralinear behavior ($\Delta < 0$) for all measurements.
Figure~\ref{fig_count} shows individual nonlinearity measurements separated by years during which the detector was used in many other experimental setups.
The first measurement of nonlinearity was performed in November 2014 using the Ortec counter (Fig.~\ref{fig_count} -- blue). 
In January 2016, we repeated the measurement---again using the Ortec counter---in order to confirm the supralinear response of the tested detector (Fig.~\ref{fig_count} -- green).
This was the most accurate measurement, with acquisition time 120~s for each individual rate measurement, repeated 20 times.
The last measurement was performed in February 2021 using the Countex counter (Fig~\ref{fig_count} -- black).
Throughout the years, we have obtained consistent results.

\section{Theoretical models of the SPAD response function}
\label{sec:models}

\subsection{Correcting the data for dead time and background}

\begin{figure*}[p]
	\centering
	\includegraphics[width=1.0\textwidth]{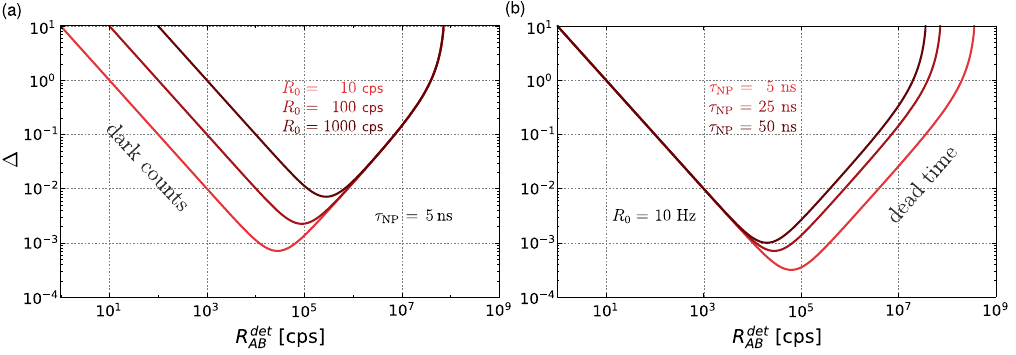}
	\caption{
		Nonlinearity $\Delta$ as a function of the detection rate $R^{\text{det}}_{\text{AB}}$ for (a) a several different values of dark counts: $R_{0}~=~10, 100, 1000$~cps ($\tau$ = const.) and (b) dead time: $\tau = 5, 25, 50~$ns ($R_{0}$ = const.). 		
	}
	\label{fig_dead_time}
\end{figure*}

There have been many SPAD response models proposed that model saturation effects and noise \cite{Muller1973,Ware2007,Karami2010,Stipcevic2013,Kornilov2014,Wang2016,Wayne2017,Straka2020}. The model used to fit the presented results considers a non-paralyzable dead time $\tau_{\text{NP}}$, dark count rate $R_0$, and neglects afterpulsing. The rate of detection events as a function of the incident rate $R$ then follows a standard formula
\begin{equation} \label{eq:model}
	f_{\text{NP}}\left(R\right) = \frac{R+R_{0}}{1+\left(R+R_{0}\right)\tau_{\text{NP}}}.
\end{equation}
In Fig.~\ref{fig_dead_time}, we illustrate the effects of dead time and dark counts, when this model is applied to nonlinearity (\ref{NL}).
The left slope represents the background noise characterized by dark counts and the right slope represents dead time saturation.
Other theoretical models shown below yield similarly V-shaped nonlinearity $\Delta$ with analogous scaling.

\begin{figure*}[p]
	\centering
	\includegraphics[width=\linewidth]{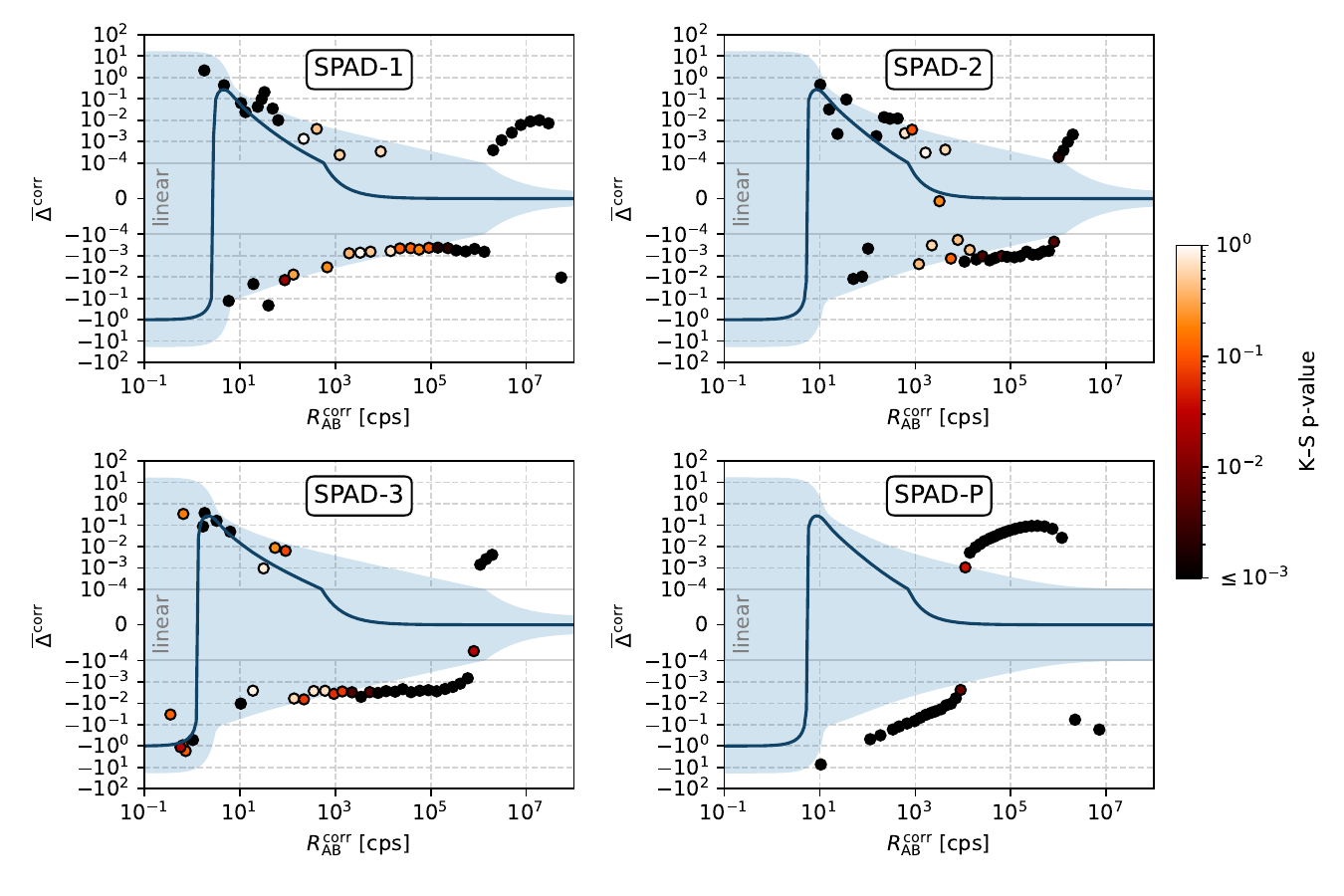}
	\caption{Measured average nonlinearities after all detection rates are corrected for dead-time and dark-count effects \eqref{eq:model_corr} based on the best-fit values given in Table~\ref{table_2} below. The points represent 30-sample averages; the solid lines are the theoretical mean values assuming the model \eqref{eq:model} holds; the blue area is a 95\% confidence interval for the average values. Point color shows the Kolmogorov--Smirnov p-value, which tests whether the 30 constituent samples come from the expected distribution. Dark points are very unlikely to occur under the presumed model.}
	\label{fig_datacorrected}
\end{figure*}

To correct for nonlinearities under this model, one can apply the formula
\begin{equation}\label{eq:model_corr}
	R^{\text{corr}}_i = \frac{R^{\text{det}}_i}{1-R^{\text{det}}_i \tau_{\text{NP}}} - R_0
\end{equation}
with $i \in \{\text{A, B, AB}\}$ to the measured detection rates. Fig.~\ref{fig_datacorrected} shows the nonlinearity values calculated from corrected rates. The same data as presented in the main text are used, and the values of $R_0$ and $\tau_{\text{NP}}$ are taken from the least-squares fits (see Fig.~\ref{fig_dead_time_2} and Table~\ref{table_2} below).

If the model \eqref{eq:model} holds, then the corrected nonlinearity should be close to zero. However, due to the Poissonian uncertainty \eqref{eq:rate_uncertainty} and the nonlinear nature of the formula \eqref{NL_app}, there is a nonzero bias (solid line in Fig.~\ref{fig_datacorrected}). The exact probability distribution of $\Delta^{\text{corr}}$ needs to be computed numerically for $R^{\text{corr}}_{\text{AB}} < 10^2$~cps, because a normal approximation no longer holds \cite{HlousekGitHub2021}.

Like in the main text, each point represents the average from 30 measurements. To test the compatibility of the corrected data and the model \eqref{eq:model}, we test both the average value and the distribution of the 30 samples for each point. The average nonlinearity is compared to a 95\% confidence interval, and the Kolmogorov--Smirnov (KS) p-value of the distribution is denoted by the point color. The p-value is a probability that, assuming the model holds, we would obtain ``worse'' data than those measured, as parameterized by the KS statistic (see Statistical Methods below). Very low p-values show that statistical errors alone are unlikely to explain the data.

Many points show a disagreement with the theoretical model, with the difference standing out more in comparison to uncorrected data (Fig.~\ref{fig_dead_time_2}), especially for $R_{\text{AB}}^{\text{det}} \gtrsim 10^6$~cps.

Accordingly, the basic model \eqref{eq:model} does not provide a satisfactory explanation of the measured data, so we should justify its use as opposed to other dead-time models, and quantify the effect of afterpulsing.

\subsection{Dead time models}

The effect of dead time in Geiger-mode SPADs is akin to Geiger-M\"{u}ller (GM) counters \cite{Levert1943,Feller1948,Albert1953,Evans1955,Takacs1958,Muller1973,Muller1988,Knoll1989,Muller1991,Gardner1997}.
Two basic types of idealized models for dead time have been defined \cite{Levert1943,Evans1955,Knoll1989}.
Namely, it is the paralyzable dead time $\tau_{\text{P}}$ model
\begin{equation} \label{eq_modelNP}
	f_{\text{P}}\left(R\right) = \left(R+R_{0}\right)\exp(-\left(R+R_{0}\right)\tau_{\text{P}}),
\end{equation}
and the non-paralyzable dead time $\tau_{\text{NP}}$ model \eqref{eq:model}.
For the non-paralyzable case, each \emph{registered} detection is followed by dead time, during which no further events are registered.
In the paralyzable case, dead time follows every photon absorption, even those that occur within a previous dead time and are not otherwise recorded. This case covers the fact that secondary detections in GM tubes still require quenching, but are not recognized due to low voltage output.

\begin{figure*}
	\centering
	\includegraphics[width=.9\textwidth]{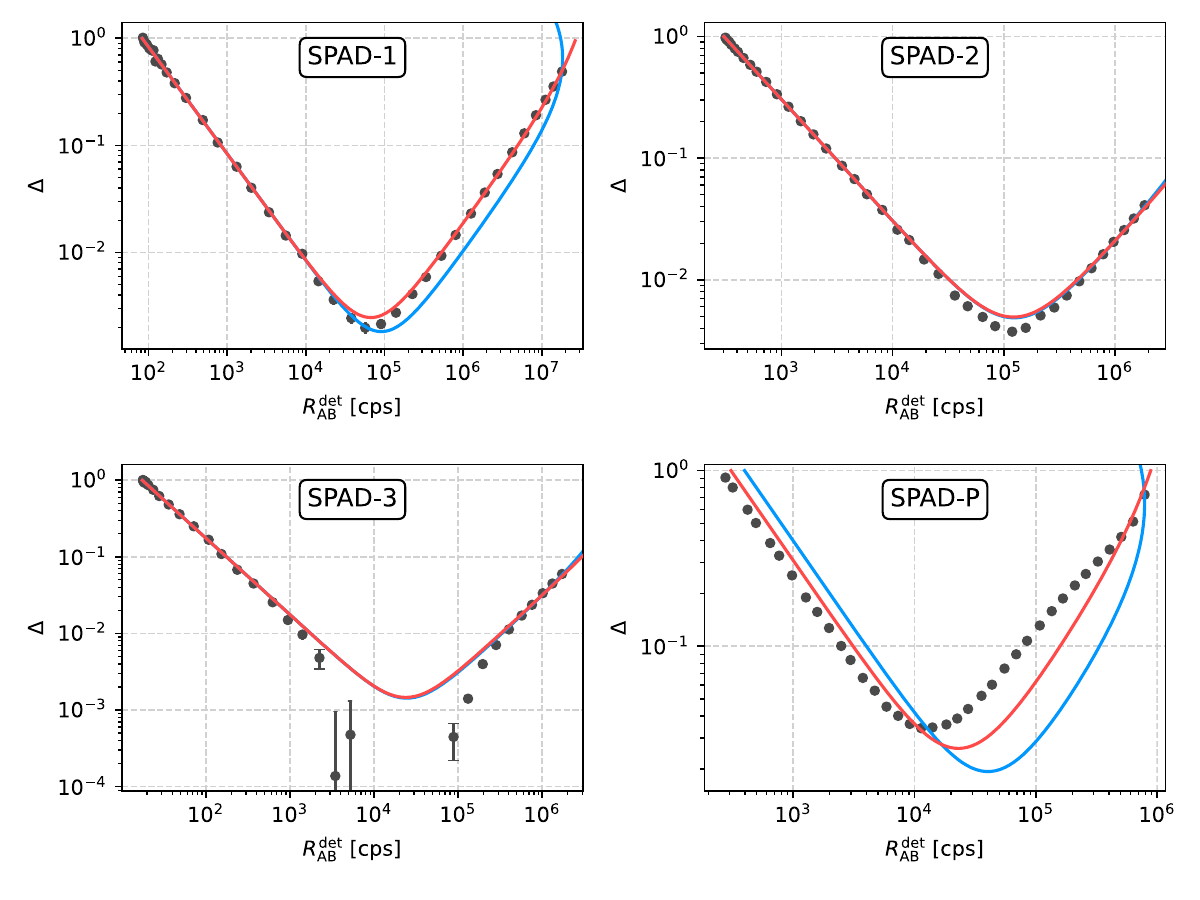}
	\caption{
		Nonlinearity data fits using two models: the $\tau_{\text{NP}}$ model (red), and the $\tau_{\text{P}}$ model (blue). Note the right-side turning point of the paralyzable model (blue), which is a consequence of non-monotonous response to illumination.
	}
	\label{fig_dead_time_2}
\end{figure*}

\begin{table*}
	\centering
	\caption{
		Comparison of directly measured dark counts $R_{0}$ and recovery times $\tau_{\text{R}}$, and dark counts and dead times $\tau_{\text{NP}}$, $\tau_{\text{P}}$, that were the best fits of individual response models. The last two columns show which scheme the hybrid models converge to.
	}
	\label{table_2}
	\begin{ruledtabular}
		\begin{tabular}{ccccccccc} 
			\multicolumn{1}{c}{\multirow{2}{*}{}} & \multicolumn{2}{c}{measured} & \multicolumn{2}{c}{$\tau_{\text{NP}}$ model} & \multicolumn{2}{c}{$\tau_{\text{P}}$ model} & \multirow{2}{*}{$\tau_{\text{NP}}$-$\tau_{\text{P}}$ model} & \multirow{2}{*}{$\tau_{\text{P}}$-$\tau_{\text{NP}}$ model}  \\\cline{2-7}
			\multicolumn{1}{c}{} & $R_{\text{0}}$ [cps] & $\tau_{\text{R}}$ [ns] & $R_{\text{0}}$ [cps] & $\tau_{\text{NP}}$ [ns] & $R_{\text{0}}$ [cps] & $\tau_{\text{P}}$ [ns] & & \\ 
			\hline
			SPAD-1 & 88(3) & 29.5(5) & 83(4) & 36.7(1) & 80(40) & 20.0(1) & $\tau_{\text{NP}}$ model & $\tau_{\text{NP}}$ model \\ 
			SPAD-2 & 314(5) & 47.0(5) & 304(2) & 40.2(4) & 304(1) & 38.9(3) & $\tau_{\text{P}}$ model & $\tau_{\text{P}}$ model \\ 
			SPAD-3 & 20(2) & 56.6(6) & 17.5(7) & 61(1) & 17.5(6) & 58.2(9) & $\tau_{\text{P}}$ model & $\tau_{\text{P}}$ model \\ 
			SPAD-P & 264(1) & 517(6) & 300(200) & 1130(20) & 400(500) & 467 & $\tau_{\text{NP}}$ model & $\tau_{\text{NP}}$ model \\ 
		\end{tabular}
	\end{ruledtabular}
\end{table*}

\begin{table}
	\centering
	\caption{
		Evaluated $\chi^2/\nu$ values of the fitted theoretical response models for tested SPADs.
	}
	\label{table_3}
	\begin{ruledtabular}
		\begin{tabular}{ccc}
			$\chi^2/\nu$	& $\tau_{\text{NP}}$ model & $\tau_{\text{P}}$ model \\ 
			\hline
			SPAD-1 & 250 & 27000 \\ 
			SPAD-2 & 12 & 8.4 \\ 
			SPAD-3 & 40 & 31 \\ 
			SPAD-P & $9.8\times10^4$ & $1.1\times10^6$ \\ 
		\end{tabular}
	\end{ruledtabular}
\end{table}

In the case of GM counters, single-parameter dead-time models are just an approximation. Hybrid models were proposed by combining paralyzable and non-paralyzable dead times \cite{Lee2000}.
There are two variants; the NP-P model
\begin{equation}
	f_{\text{NP-P}}\left(R\right) = \frac{\left(R+R_{0}\right)\exp(-\left(R+R_{0}\right)\tau_{\text{P}})}{1+\left(R+R_{0}\right)\tau_{\text{NP}}},
\end{equation}
and the P-NP model \cite{Muller1991,Lee2009}
\begin{equation}
	f_{\text{P-NP}}\left(R\right) = \frac{\left(R+R_{0}\right)\exp(-\left(R+R_{0}\right)\tau_{\text{P}})}{1+\left(R+R_{0}\right)\tau_{\text{NP}}\exp(-\left(R+R_{0}\right)\tau_{\text{P}})}.
\end{equation}

We have tested whether these theoretical models can be used empirically to fit the measured nonlinearity data. We found that all hybrid-model fits converge to either the paralyzable or non-paralyzable case. Fig.~\ref{fig_dead_time_2} shows both of these response models applied to the measured nonlinearity of all detectors. A complete list of the measured parameters and best-fit parameters is given in Table~\ref{table_2}.

As in the main text, we evaluated $\chi^2$ to demonstrate that investigated models significantly deviate from the measured nonlinearity. In Table~\ref{table_3}, we show the chi-squared per one degree of freedom, $\chi^2/\nu$, where $\nu$ is the number of data points minus the number of fitting parameters.

Both the $\chi^2$ values and the plots in Fig.~\ref{fig_dead_time_2} show that the paralyzable and hybrid models do not offer a better explanation of the data; nor can they satisfactorily match the trend in the middle of the graphs of SPAD-1--3, where all nonlinearities seem to be systematically lower.

\subsection{Afterpulsing}

Actively quenched SPADs exhibit afterpulsing and twilight pulsing that affect the mean detection rate \cite{Straka2020}. Both effects can be evaluated numerically \cite{Straka2020Code}, or -- if we neglect the temporal distribution of afterpulses -- an approximate rate formula can be used \cite{Straka2020},
\begin{equation}\label{eq:modelAP}
	f_{\mathrm{AP}}(R) = \left( \left[ \frac{1}{(R+R_0)} - \alpha \right] e^{-\langle n_{\mathrm{AP}}\rangle} + \tau_{\text{NP}} \right)^{-1}.
\end{equation}

The new parameters are the mean number of afterpulses per detection $\langle n_{\mathrm{AP}}\rangle$, and the twilight-pulse proportionality constant $\alpha$ that introduces rate dependence. As one would expect, when both of these parameters are zero, the formula \eqref{eq:modelAP} is reduced to the basic non-paralyzable model \eqref{eq:model}.

The key observation here is parameter degeneracy. In the formula \eqref{eq:modelAP}, the effect of afterpulsing can be substituted by adjusting the dead time and dark count variables in \eqref{eq:model}. If we put the model parameters in square brackets, the equivalence can be expressed as
\begin{equation}\label{eq:params_equivalence}
	\begin{split}
		f_{\text{AP}} \Big( R \Big) \Big[R_0,\tau_{\text{NP}},\langle n_{\text{AP}} \rangle, \alpha \Big]\\
		\equiv f_{\text{NP}} \Big( R/\mathrm{e}^{-\langle n_{\mathrm{AP}}\rangle} \Big) \Big[ R_0/\mathrm{e}^{-\langle n_{\mathrm{AP}}\rangle}, \tau_{\text{NP}} - \alpha\mathrm{e}^{-\langle n_{\mathrm{AP}}\rangle} \Big].
	\end{split}
\end{equation}

Let us now substitute both rate models into nonlinearity (\ref{NL}), assuming balanced splitting $R^{\text{det}}_{\mathrm{A}} = R^{\text{det}}_{\mathrm{B}}$ and the parameter equivalence \eqref{eq:params_equivalence},
\begin{equation}\label{eq:nonlinearity}
	\Delta = \frac{2 R^{\text{det}}_{\mathrm{A}}}{R^{\text{det}}_{\mathrm{AB}}} - 1 = \frac{2f[\frac{1}{2}f^{-1}(R^{\text{det}}_{\mathrm{AB}})]}{R^{\text{det}}_{\mathrm{AB}}} - 1.
\end{equation}
One finds that $\Delta_{\text{AP}} \equiv \Delta_{\text{NP}}$, which follows from the equivalence \eqref{eq:params_equivalence} and the scaling of $R$ by a linear factor there. The principle is that both $f_{\text{AP}}^{-1}$ and $f_{\text{NP}}^{-1}$ yield incident rates that differ by the same linear factor, and their ratio does not change upon the multiplication by 1/2. A subsequent application of $f$ therefore gives identical rates for both models. As a result, all fitted models are identical and yield the same $\chi^2$ either with or without afterpulsing.

\begin{figure}
	\centering
	\includegraphics[width=\linewidth]{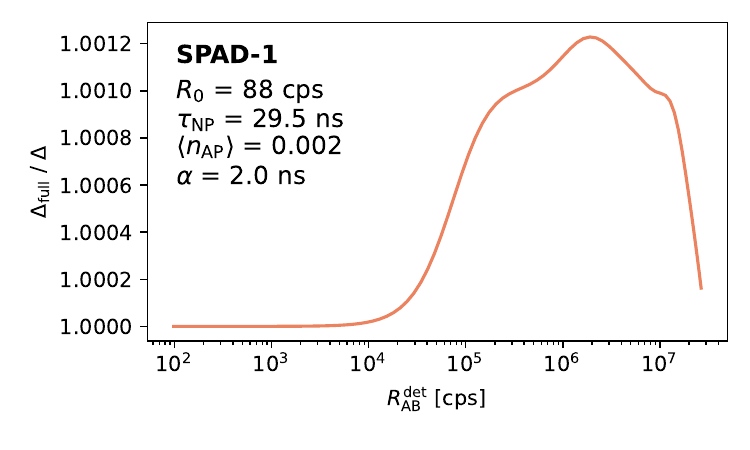}
	\caption{The full afterpulsing model is compared to the simple model \eqref{eq:modelAP}, and thus to an equivalent basic model \eqref{eq:model} using \eqref{eq:params_equivalence}. The difference is shown for SPAD-1. It corresponds to a relative change of the fits depicted in Fig.~\ref{fig_dead_time_2} if a full afterpulsing correction was made.}
	\label{fig:afterpulsing}
\end{figure}

A further consideration would be a more complex afterpulsing model that considers the temporal distribution of afterpulses \cite{Straka2020}. This model is based on computationally demanding Monte-Carlo simulations and is therefore unsuitable for least-squares fitting. However, the difference between this full model $\Delta_{\text{full}}$ and the simple model \eqref{eq:modelAP} can be evaluated for an example case. In Fig.~\ref{fig:afterpulsing}, the difference is shown for SPAD-1 based on the afterpulsing data measured in Ref.~\onlinecite{Straka2020}. Note that we have already established the equivalency of \eqref{eq:model} and \eqref{eq:modelAP}, so we can conclude that the full model introduces a very small correction to the basic model we used to fit the data.

\subsection{Discussion of the models}

An example case of an actively quenched SPAD, represented by SPAD-1, shows that an afterpulsing correction to our nonlinearity model would yield a relative difference of $\approx 0.1\%$ in $\Delta$. A~slightly simplified afterpulsing correction that can be used for data fitting yields no difference from the basic model. For this reason, we can conclude that afterpulses do not play a role in fitting the nonlinearity data of actively quenched SPADs.

The other detectors exhibit a much more complex behavior. Passively quenched SPADs exhibit paralyzable dead time, but also a rising efficiency curve after each detection \cite{Cova1996Apr}, which makes the model too complex to be parameterized by a few measurable quantities.

SNSPDs show no afterpulsing, but, to our knowledge, their (non-)paralyzability has not been confirmed. Additionally, both SNSPDs and passively quenched SPADs exhibit changing efficiency after each detection, an effect that is both non-negligible and complex. In the case of SNSPDs, the single-photon and two-photon efficiencies also depend on the bias current. For these reasons, no standard rate model has been formulated that would be sufficiently accurate.

Upon exploration of certain empirical models combining non-paralyzable and paralyzable dead time, we found that none of them offer a significant advantage over the basic non-paralyzable model. SPAD-1--3 are non-paralyzable. SPAD-2 and SPAD-3 are better fitted with the paralyzable model, but the difference is very small, as can be seen from Fig.~\ref{fig_dead_time_2}. SPAD-P is significantly better fitted by the non-paralyzable model. Consequently, we use a single model \eqref{eq:model} for all SPADs in the manuscript.

\subsection{Ad hoc hypothesis}

The most significant supralinearity ($\Delta < 0$) among SPADs is exhibited by SPAD-3. Here we use an ad hoc model to fit the data and assess the feasibility of various factors as possible explanations of the supralinearity.

The data can be fit with a modified version of formula \eqref{eq:model},
\begin{equation}\label{eq:adhoc_model}
	f(\Phi) = \frac{1}{(a_0 + a_1 \Phi + a_2 (\Phi/\Phi_1)^b)^{-1}+\tau},
\end{equation}
where the term $a_0$ corresponds to a constant dark count rate, $a_1$ corresponds to a constant detection efficiency, and $\Phi$ is the incident photon flux. The last term (with $b$ = 1.004 and $R_1$ = $10^6$~cps) is empirical and can be attributed to either of these parameters as a hypothetical rate-dependent term.

\begin{figure}
	\centering
	\includegraphics{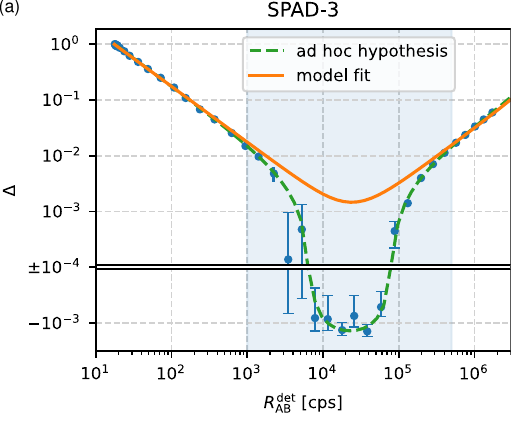}

	\vspace{2mm}

	\includegraphics{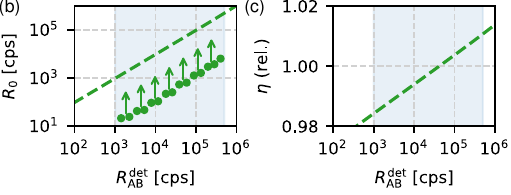}

	\caption{Depiction of the hypothetical rate dependencies of dark counts and detection efficiency that would be necessary to account for the suprealinearity of SPAD-3 in the highlighted region. (A) depiction of an ad hoc model; followed by respective attribution of the extra empirical term to dark counts (B) and efficiency (C). The points in B represent the minimal values calculated from the data points.}
	\label{fig:adhoc}
\end{figure}

Fig.~\ref{fig:adhoc} depicts the hypothesis (A), and the corresponding dependence of dark counts (B) and efficiency (C) that would be required to match the hypothesis. The dark count rate dependence is arbitrary up to a linear factor $a \Phi$, but we can calculate the absolute minimum dark count rate directly from the data (green points in Fig.~\ref{fig:adhoc}B).

Dependence on the dark count rate was ruled out by an independent pulsed measurement, where the count rate can be inferred from detections registered outside the pulse window. The dark count rate was found to be below $30$~cps for a signal of $10^5$~cps mean detection rate, whereas the nonlinearity data would require at least $R_0(10^5~\text{cps})$ > 1000~cps. This means that rate-dependent dark count rate is not the cause behind the supralinearity of SPAD-3.

If we attribute the supralinearity to rate-dependent efficiency, we are left with relative changes depicted in Fig.~\ref{fig:adhoc}C. While efficiency decreasing with rate has recently been reported in Ref.~\onlinecite{Raupach2022Apr} for InGaAs SPADs subjected to pulsed signals, here the efficiency would need to increase. For a SPAD, the efficiency increases with bias voltage, which should be fixed in its steady state by a DC voltage supply. However, we need to rule out any temporal dependence of the bias voltage, which is addressed in the next section.

\subsection{Efficiency settling}
\begin{figure}
	\centering
	\includegraphics[width=.9\linewidth]{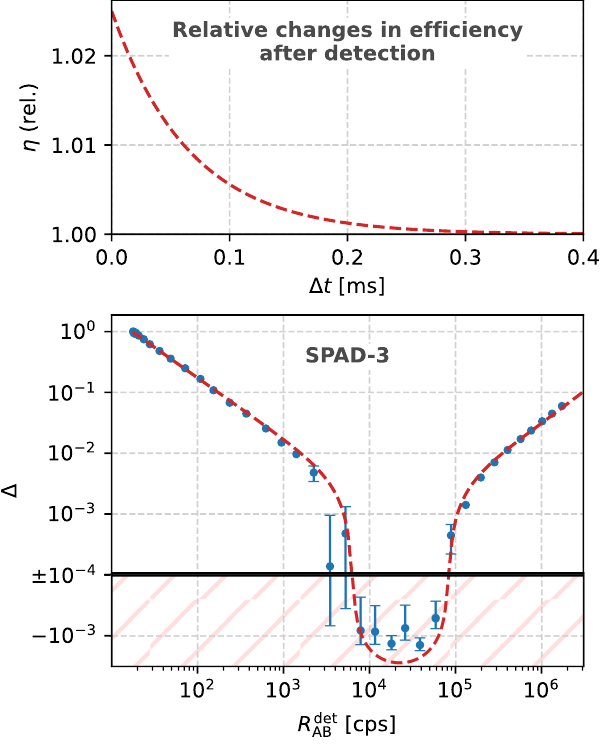}
	\caption{Estimation of the necessary magnitude of the efficiency settling effect. If the detection efficiency $\eta$ settled as depicted in the top graph, the measured supralinearity could be approximately achieved.}
	\label{fig:settling_bias}
\end{figure}
A possible explanation for the supra-linear response of a SPAD would be time-dependent efficiency after each detection. When an actively quenched detector resets, the overall response may fluctuate \cite{Ware2007}. In order for a detector to exhibit supra-linearity, the dominant effect would have to include decreasing efficiency over time after each detection. The slope of this decrease needs to be significant on time scales close to inverse mean count rates where supra-linearity occurs ($\sim$10--100~\textmu{}s). This could, in principle, be caused by a settling bias voltage, which affects detection efficiency.

In an attempt to roughly estimate the necessary magnitude of such an effect, we modeled the efficiency with an exponential. The results are shown in Fig.~\ref{fig:settling_bias}. To achieve supralinearity roughly similar to the data observed on SPAD-3, the relative decrease in efficiency would have to be $2.5\%$ over more than 0.1~ms. However, as the quenching electronics operate on a GHz time scale, such a slow and significant effect is extremely unlikely.

This effect can be ruled out based on interarrival histograms of detections under continuous illumination. For constant efficiency, interarrival times follow an exponential distribution, save for dead time (here 57 ns) and afterpulsing effects that take place on time scales <$10$~\textmu{}s. For time-dependent efficiency, the interarrival time $\Delta t$ is distributed with the probability density
\begin{equation}
	p(\Delta t) \approx R \eta(\Delta t) \exp \left( -R \int_0^{\Delta t} \eta(t') \mathrm{d}t' \right),
\end{equation}
where $R$ is the incident rate and $\eta$ is a relative term describing the changes in detection efficiency. Interarrival histograms then sample this distribution.

Let us assume the hypothesis shown in Fig.~\ref{fig:settling_bias} that $\eta(\Delta t) = 1 + 0.025 \exp(-15 \Delta t \times \mathrm{ms}^{-1})$. Then, for $\Delta t \gtrsim 0.2$~ms, $\eta(\Delta t) \approx 1$ and the distribution will scale as an exponential---$p(\Delta t) \propto \exp(-R \Delta t)$---which is the same scaling as for constant efficiency. On shorter time scale, $\Delta t \lesssim 0.2$~ms, the distribution will be more complex.

\begin{figure}
	\centering
	\includegraphics[width=.9\linewidth]{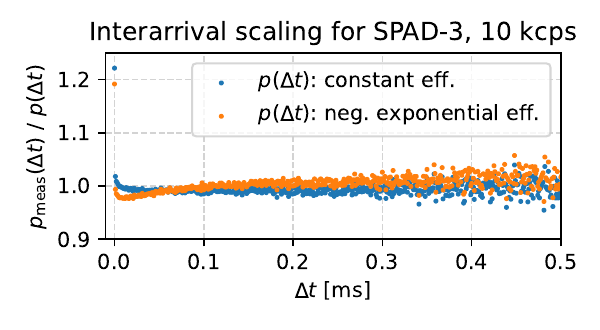}
	\caption{Scaling of interarrival times for CW illumination at 10~kcps for SPAD-3. The data are taken relative to expected histogram values for constant efficiency and settling efficiency as depicted in Fig.~\ref{fig:settling_bias}. One point corresponds to a 1~\textmu{}s window. The peak near zero is due to afterpulsing. The model of constant efficiency shows consistent ratio of its expected values relative to the data, as opposed to settling efficiency, where the scaling is different below $\Delta t < 0.2$~ms.}
	\label{fig:IAhist}
\end{figure}

Data in Fig.~\ref{fig:IAhist} show the measured interarrival histograms relative to the two models---constant and settling efficiency. Because the difference is small, the histogram values $p_\text{meas}(\Delta t)$ are divided by the expected values $p(\Delta t)$ to visually distinguish the two cases more clearly. A horizontal trend means that the scaling corresponds to the expectation up to a constant proportionality factor. It can be seen that the data follow a negative exponential for constant efficiency, whereas the comparison to settling efficiency does not result in a horizontal data trend for $\Delta t < 0.2$~ms. This provides good evidence that the efficiency settling effect is not the reason for supralinearity of SPAD-3.

\subsection{Suprealinearity summary}
Here we summarize the factors that we rule out as causes of supralinearity, as demonstrated on SPAD-3.

\paragraph{Dead time} Hybrid dead-time models do not satisfactorily fit the data. The non-paralyzable nature of dead time (more strictly, recovery time) is directly measurable, and has been observed to slightly increase with count rate \cite{Straka2020}, which actually works against supralinearity. To match the data, dead time values would have to reach unphysical negative values.

\paragraph{Dark counts} Dark counts are known to fluctuate \cite{Karami2010}, but for nonlinearity, systematic dependence of the mean dark count rate on the incident rate is needed. This dependence would need to span orders of magnitude, which was ruled out by a pulsed measurement.

\paragraph{Efficiency} Detection efficiency depends on the SPAD bias voltage, which is held constant by an active quenching circuit. There could be, in theory, a temporal dependence after each detection. However, the settling effect necessary to explain supralinearity between $10^4$--$10^5$~cps would have to be both significant and low-frequency, which was ruled out by time-resolved measurements.

\paragraph{Afterpulsing} The probability of an afterpulse has a constant and linear contribution for CW signals \cite{Wayne2017,Straka2020}. This was shown to have no effect on nonlinearity, and the effect of the afterpulse temporal distribution is negligible, as we demonstrated.

The above considerations are based on current knowledge of SPAD operation, and none of them offer an explanation for the supralinearity that we observed. 

\section{Statistical methods}
This section elaborates on the statistical methods used to analyze and fit the SPAD data.

Due to the integration time $T$ = 20 s, all the measured numbers of counts are $R^{\text{det}} T > 300$, which is sufficient for the normal approximation of the Poisson distribution. We therefore assume that all acquired count rates are normally distributed $\mathcal{N}(\theta, \sigma^2)$ with variance $\sigma^2$ \eqref{eq:rate_uncertainty} as a function of a mean-value parameter $\theta$,
\begin{equation}\label{eq:Rdet_normalDist}
	R^{\text{det}} \sim \mathcal{N} \left( \theta, (1-\theta\tau_{\text{NP}})^2 \theta/T \right).
\end{equation}
We also assume that nonlinearity $\Delta$, as a function of these rates, is also normally distributed, which requires the standard deviations to be small enough with respect to the mean values. The estimation of $\Delta$ is then carried out based on $N = 30$ samples. Due to the normality of the data, a least-squares fit of the model can be performed (Fig.~\ref{fig_dead_time_2}).

However, when applying the correction \eqref{eq:model_corr} on the data, the normality of $\Delta$ ceases to hold. This happens when the detection rate approaches the background rate, as then the corrected value fluctuates significantly compared to the mean value. For such low rates, dead time correction can be neglected and $R^{\text{corr}} \approx R^{\text{det}} - R_0$,
\begin{equation} \label{eq:Rcorr_PDFlow}
	R^{\text{corr}} \sim \mathcal{N} \left( R, (R + R_0)/T \right) \quad \text{for} \quad R \ll 1/\tau_{\text{NP}},
\end{equation}
where $R$ denotes the incident rate. Due to Poissonian fluctuations of the dark counts, the corrected values can also end up negative. The probability density of
\begin{equation}\label{eq:Delta_corr}
	\Delta^{\text{corr}} = \frac{R^{\text{corr}}_{\text{A}} + R^{\text{corr}}_{\text{B}}}{R^{\text{corr}}_{\text{AB}}} - 1
\end{equation}
can be computed from \eqref{eq:Rcorr_PDFlow} by using the probability-density transformations
\begin{align} \label{eq:PDF_convolution}
	p_{\text{(Z=X+Y)}}(z) &= \int_{-\infty}^{\infty} p_{\text{X}}(t) p_{\text{Y}}(z-t) \mathrm{d}t, \\
	p_{\text{(Z=X/Y)}}(z) &= \int_{-\infty}^{\infty} p_{\text{X}}(zt) p_{\text{Y}}(t) |t| \mathrm{d}t.
\end{align}

For $R \to 0$, the distribution of $\Delta^{\text{corr}}$ tends towards the Lorentz distribution with undefined variance, so that one cannot employ the central limit theorem to arrive at estimations of $N$-sample averages. As per \eqref{eq:PDF_convolution}, the probability density of $\sum_{i=1}^N \Delta^{\text{corr}}_i$ is computed numerically by $(N-1)$ successive convolutions. This allows plotting the mean value and confidence intervals shown in Fig.~\ref{fig_datacorrected}, as well as evaluating the Kolmogorov--Smirnov statistic.

For $R \gg 0$, the distribution of $\Delta^{\text{corr}}$ can be regarded as normal, with its standard deviation obtained by locally linear propagation of \eqref{eq:Rdet_normalDist}. However, as we are concerned with small values of $\Delta$, we need to account for a bias in the mean value.

We note that all the measured rates in \eqref{eq:Delta_corr} are independent, and the scheme is balanced, $\left\langle R^{\text{corr}}_{\text{A}} \right\rangle = \left\langle R^{\text{corr}}_{\text{B}} \right\rangle$, and so
\begin{equation}
	\langle \Delta^{\text{corr}} \rangle = 2\left\langle R^{\text{corr}}_{\text{A}} \right\rangle \left\langle \frac{1}{R^{\text{corr}}_{\text{AB}}} \right\rangle - 1.
\end{equation}
We employ the function \eqref{eq:model_corr}, where $R^{\text{det}}_i$ is a normal-distributed variable described by \eqref{eq:Rdet_normalDist}. The mean values $\theta_i$ for $i$ = A,~AB are given by the function \eqref{eq:model}, and the incident rate is $R$ for $i$ = AB and $R/2$ for $i$ = A. The main source of the bias is nonlinear dependence of the averaging terms on the random variables. We expand the averaging terms in a Taylor series,
\begin{equation}
	f(R^{\text{det}}_i) \approx \sum_{n=0}^2 \left(\frac{\partial^n f}{\partial (R^{\text{det}}_i)^n} \right) (\theta_i) \frac{(R^{\text{det}}_i - \theta_i)^n}{n!},
\end{equation}
where it is sufficient to stop at the quadratic term. Averaging over $R^{\text{det}}_i \sim \mathcal{N}(\theta_i, \sigma_i^2)$ yields non-zero only for even $n = 2k$:
\begin{equation}
	\left\langle \frac{(R^{\text{det}}_i - \theta_i)^{2k}}{(2k)!} \right\rangle = \frac{\left( \sigma_i^2/2 \right)^k}{k!}.
\end{equation}

We can also safely assume a short dead time and low background --- $\tau \ll T$ and $\tau R_0 \ll 1$. This all results in
\begin{equation}\label{eq:Delta_corr_bias}
	\langle \Delta^{\text{corr}} \rangle \approx \frac{R + R_0}{T R^2},
\end{equation}
which holds well for $R \gtrsim R_0$.

The above methods allow us to establish the probability distribution of $\overline{\Delta}^{\text{corr}} = \frac{1}{N} \sum_{i=1}^{N} \Delta^{\text{corr}}_i$ for the whole range of incident rates $R$. Then, one way of testing the validity of the data is the confidence interval of $\overline{\Delta}^{\text{corr}}$. A second test evaluates whether the samples $\Delta^{\text{corr}}$ conform to the expected distribution. However, due to non-normality, it would be difficult to evaluate measures based on multivariate probability density akin to $\chi^2$. Instead, we compare cumulative distributions.

The Kolmogorov--Smirnov statistic $D_N$ is defined as the maximum difference between the empirical and theoretical cumulative distributions. For a certain $R^{\text{det}}_{\text{AB}}$, let the number of samples of $\Delta^{\text{corr}}$ lower than $\Delta$ be $n(\Delta)$. Then, the empirical cumulative distribution is  $S_N(\Delta) = n(\Delta)/N$ with $S_N \in [0,1]$. Let the theoretical cumulative distribution be $S(\Delta) = \Pr[\Delta^{\text{corr}} < \Delta]$. We compare them by defining the statistic
\begin{equation}\label{eq:Dn}
	D_N := \sup_\Delta \big| S_N(\Delta) - S(\Delta) \big|.
\end{equation}
Then, for many samples, as $N \to \infty$, the quantity ($\sqrt{N} D_N$) follows a known distribution assuming $S(\Delta)$ holds \cite{Feller1948June}. This quantity can therefore be evaluated for each set of $N$ measurements, as depicted in Fig.~\ref{fig_datacorrected}. The p-value is then the probability $\Pr[D_N > D^{\text{meas}}_N]$, which tests the null hypothesis that the data conform to the theoretical model.

\bibliography{NL_GLOBAL_bibliography}

\end{document}